%% file: main.tex
\documentclass[11pt,a4paper]{article}

\usepackage[utf8]{inputenc}
\usepackage[T1]{fontenc}
\usepackage{lmodern}
\usepackage{amsmath,amssymb}
\usepackage{graphicx}
\usepackage{booktabs}
\usepackage[tableposition=top]{caption}
\usepackage{siunitx}
\usepackage{xcolor}
\usepackage[margin=2.5cm]{geometry}
\usepackage[colorlinks=true,linkcolor=blue!60!black,citecolor=blue!60!black,urlcolor=blue!60!black]{hyperref}
\usepackage[numbers,sort&compress]{natbib}
\usepackage{authblk}

\input{macros}

\input{q/numbers}
\newif\ifextrasec\extrasecfalse   
\newif\ifsuperchicsec\superchicsecfalse   
\makeatletter
\renewcommand{\subsection}{\@ifstar{\ss@nosubsection}{\ss@nosubsection}}
\newcommand{\ss@nosubsection}[2][]{\par}
\makeatother

\title{Probing the $\tau$ anomalous magnetic and electric dipole moments\\
       in photon-induced $\gamma\gamma\to\tau^+\tau^-$ production at the FCC-hh}
\author[1]{A.~Senol\thanks{\href{mailto:senol_a@ibu.edu.tr}{senol\_a@ibu.edu.tr}}}
\author[1]{H.~Denizli\thanks{\href{mailto:denizli_h@ibu.edu.tr}{denizli\_h@ibu.edu.tr}}}
\author[2,3]{M.~K{\"o}ksal\thanks{\href{mailto:mkoksal@cumhuriyet.edu.tr}{mkoksal@cumhuriyet.edu.tr}}}
\affil[1]{Department of Physics, Bolu Abant Izzet Baysal University, 14280, Bolu, T\"{u}rkiye}
\affil[2]{Department of Physics, Sivas Cumhuriyet University, 58140, Sivas, T\"{u}rkiye}
\affil[3]{Sivas Cumhuriyet University Nanophotonics Application and Research Center (CUNAM), 58140, Sivas, T\"{u}rkiye}

\date{\today}

\begin{document}
\maketitle

\begin{abstract}
\input{q/sections/abstract}
\end{abstract}

\input{q/sections/introduction}
\input{q/sections/theory}
\input{q/sections/simulation}
\input{q/sections/selection}
\input{q/sections/statistics}
\input{q/sections/results}
\input{q/sections/systematics}
\input{q/sections/conclusions}
\input{q/sections/acknowledgements}

\bibliographystyle{unsrtnat}
\bibliography{refs}

\end{document}

%% file: macros.tex
\newcommand{\atau}{a_\tau}
\newcommand{\dtau}{d_\tau}
\newcommand{\ka}{\kappa_a}
\newcommand{\kta}{\tilde\kappa_a}
\newcommand{\mtt}{m_{\tau\tau}}
\newcommand{\mttot}{m_{\mathrm{T}}^{\mathrm{tot}}}
\newcommand{\tauh}{\tau_{\mathrm{h}}}
\newcommand{\gaga}{\gamma\gamma}
\newcommand{\ecm}{e\,\mathrm{cm}}
\newcommand{\invab}{\mathrm{ab}^{-1}}

\newcommand{\Stwo}{S^2}
\newcommand{\MG}{\textsc{MadGraph5\_\allowbreak aMC@NLO}}
\newcommand{\Pythia}{\textsc{Pythia}\,8}
\newcommand{\Delphes}{\textsc{Delphes}}
\newcommand{\SuperChic}{\textsc{SuperChic}}

%% file: q/sections/abstract.tex
The sensitivity of the FCC-hh to the anomalous magnetic moment $\atau$ and the electric dipole moment $\dtau$ of the
$\tau$ lepton is studied in photon-induced $\gaga\to\tau^+\tau^-$ production with both $\tau$ leptons decaying
hadronically. Central exclusive (elastic) signal events, in which both protons remain intact, are generated with \MG\
using a \textsc{FeynRules} UFO model of the anomalous $\gamma\tau\tau$ vertex merged with the TauDecay model, so that
the $\tau$ decays and their spin correlations are included in the matrix element. Single- and double-dissociative
production is estimated with \SuperChic\ and added to the elastic signal with its own coupling dependence and with
normalisation and shape uncertainties of 25\%. \Pythia\ decays the unstable hadrons in the signal, decays the $\tau$
leptons and simulates the parton shower and hadronisation in the backgrounds, and hadronises the dissociated protons.
Signal and backgrounds are passed through the \Delphes\ simulation of the FCC-hh detector. Hadronic $\tau$ decays are
reconstructed from particle-flow tracks and photons with an algorithm based on the hadrons-plus-strips approach, with
an efficiency of \res{reco-eff}\% within the acceptance. Requiring no additional tracks from the $\tau$ vertices
retains part of the dissociative contribution and reduces the Drell--Yan and $\gaga\to WW$ backgrounds to
\res{y-bkg-rel-tot}\% of the SM yield. The couplings are constrained from the shape of the total transverse mass
distribution. For a centre-of-mass energy of 84~TeV and $30~\invab$, the expected 95\% CL limits are
$\res{ataulo-84-30}\times10^{-4} < \atau < \res{atauhi-84-30}\times10^{-4}$ and $|\dtau| <
\res{dtau-84-30}\times10^{-18}\,\ecm$, about an order of magnitude beyond the current bounds. Results are also given
for FCC-hh scenarios between 72 and 120~TeV and for an optimistic detector parametrisation.

%% file: q/sections/introduction.tex
\section{Introduction}\label{sec:intro}

The anomalous magnetic moment $\atau = (g_\tau-2)/2$ and the electric dipole moment $\dtau$ of the $\tau$ lepton are
among the least precisely known properties of the charged leptons. In the Standard Model (SM), $\atau$ is predicted
with high precision, $\atau^{\rm SM} = 1.17721(5)\times10^{-3}$ \cite{Eidelman:2007sb}, whereas $\dtau$ is
negligibly small. Therefore, a non-zero $\dtau$ would point to a new source of CP violation. New physics at a scale
$\Lambda$ typically contributes to the dipole moments of a lepton $\ell$ in proportion to $m_\ell^2/\Lambda^2$ which
favours the $\tau$ over the muon by a factor $(m_\tau/m_\mu)^2\approx 280$. The short $\tau$ lifetime rules out the spin-precession methods used for electrons and muons, apart from proposals
using bent crystals~\cite{Fomin:2018ybj}, and the $\gamma\tau\tau$ vertex is studied in production processes
instead.

Experimentally, the $\gamma\tau\tau$ vertex has been studied in $e^+e^-\to e^+e^-\tau^+\tau^-$ at LEP
\cite{DELPHI:2003nah}, in $e^+e^-\to\tau^+\tau^-$ at Belle \cite{Belle:2021ybo}, and in
photon-induced $\tau$-pair production at the LHC, in ultraperipheral lead-lead collisions
\cite{ATLAS:2022ryk,CMS:2022arf,CMS:HIN-24-011} and in proton-proton collisions. The
currently most precise constraints on photon-induced production come from CMS in proton-proton collisions at
13~TeV~\cite{CMS:2024skm}, $\atau = 0.0009^{+0.0032}_{-0.0031}$ and $|\dtau| < 2.9\times10^{-17}\,\ecm$ at 95\% CL. The
anomalous contributions grow with the invariant mass of the $\tau$ pair, so the reach improves with the available
photon-photon energy. Photon-induced production in proton-proton collisions
receives large contributions from events in which one or both protons dissociate, and a reliable interpretation
requires a consistent treatment of these contributions and of the soft survival
probability~\cite{Harland-Lang:2024zpn}. The use of $\gaga\to\tau^+\tau^-$ in proton-proton collisions to constrain the $\tau$ dipole moments was proposed
for the LHC in Ref.~\cite{Atag:2010ja}, and a measurement strategy for LHC proton collisions has been developed
recently~\cite{Beresford:2024dsc}. Sensitivity studies exist for ultraperipheral heavy-ion collisions at the
LHC~\cite{Beresford:2019gww,Dyndal:2020yen} and at the FCC-hh~\cite{Inan:2026byu}, for photon-proton collisions at
the LHC~\cite{Koksal:2017nmy}, for electron-proton colliders~\cite{Koksal:2018xyi,Gutierrez-Rodriguez:2022eyn}, and
for $e^+e^-$ and muon colliders~\cite{Billur:2013rva,Koksal:2018env,Chen:2018cxt,Denizli:2024uwv,Shao:2025xwp}. Optimal observables for the dipole moments are also discussed in Ref.~\cite{Bhide:2024nje}.

The Future Circular Collider in its hadron-hadron configuration (FCC-hh)~\cite{FCC:2018vvp,FCC:2025uan} will provide
proton-proton collisions at centre-of-mass energies of order 100~TeV and integrated luminosities of several tens of
$\invab$. The corresponding photon-photon luminosity~\cite{dEnterria:2019jty,Shao:2022cly,dEnterria:2025ecx}
extends to invariant masses of several TeV where the
anomalous contributions dominate. Several operating scenarios with $\sqrt{s}$ between 72 and 120~TeV, corresponding to dipole fields between 12 and
20~T, are currently under discussion~\cite{Zimmermann:2024fcchh,FCC:2025jtd,FCC:2025uan,FCC:2025lpp} and are listed in Table~\ref{tab:scenarios}.

\begin{table}[tbp]
  \centering
  \small
  \caption{FCC-hh operating scenarios considered in this paper, from Ref.~\cite{Zimmermann:2024fcchh}: centre-of-mass
    energy, dipole field, integrated luminosity per year and for ten years, initial number of pile-up interactions
    per bunch crossing $\mu$, and rms length of the luminous region $\sigma_z$. The values of the low- and
    high-current options (F12LL and F12HL) are also given F12.}
  \input{q/tables/scenarios}

  \label{tab:scenarios}
\end{table}

Here we estimate the FCC-hh sensitivity to $\atau$ and $\dtau$ in photon-induced $\gaga\to\tau^+\tau^-$
production for four of these scenarios, using events in which both $\tau$ leptons decay hadronically. The signal
comprises the central exclusive (elastic) process $pp\to p\,\tau^+\tau^-p$, in which both protons remain intact,
and the single- and double-dissociative processes, in which one or both protons break up. For the elastic process
the $\tau$ decays are generated inside the matrix element, which keeps the spin correlations, and the coupling
dependence is obtained by reweighting one sample. The dissociative contributions, which are absent from the \MG\
simulation, are added with \SuperChic~\cite{Harland-Lang:2020veo}, since a large fraction of them passes the
requirement of no additional tracks used to select the signal.

The paper is organised as follows. Section~\ref{sec:theory} introduces the anomalous couplings, and
Section~\ref{sec:simulation} describes the simulation of signal, dissociative contributions and backgrounds. The
reconstruction and event selection are presented in Section~\ref{sec:selection} and the statistical treatment in
Section~\ref{sec:statistics}. 
The results including the systematic uncertainties and limitations are given in Section~\ref{sec:results},
and conclusions are drawn in Section~\ref{sec:conclusions}.

%% file: q/tables/scenarios.tex
\begin{tabular}{lcccccc}
\toprule
Scenario & $\sqrt{s}$ [TeV] & dipole field [T] & $\mathcal{L}$/year [fb$^{-1}$] & $\mathcal{L}_{10\,\rm yr}$ [ab$^{-1}$] & pile-up $\mu$ & $\sigma_z$ [mm] \\
\midrule
F12 & 72 & 12 & 950 / 2000 & 9.5 / 20 & 580 / 2820 & 57 \\
F14 & 84 & 14 & 920 & 9.2 & 590 & 57 \\
F17 & 102 & 17 & 920 & 9.2 & 732 & 57 \\
F20 & 120 & 20 & 370 & 3.7 & 141 & 64 \\
\bottomrule
\end{tabular}

%% file: q/sections/theory.tex
\section{Anomalous $\gamma\tau\tau$ couplings}\label{sec:theory}

The general Lorentz-invariant $\gamma\tau\tau$ vertex for on-shell $\tau$ leptons can be written in terms of
the Dirac, Pauli and electric-dipole form factors $F_1(q^2)$, $F_2(q^2)$ and $F_3(q^2)$ as
\begin{equation}
  \Gamma^\mu \;=\; -ie\left[F_1(q^2)\,\gamma^\mu + F_2(q^2)\,\frac{i\,\sigma^{\mu\nu}q_\nu}{2m_\tau}
                  + F_3(q^2)\,\frac{\sigma^{\mu\nu}q_\nu\gamma^5}{2m_\tau}\right] ,
  \label{eq:vertex}
\end{equation}
where $\sigma^{\mu\nu}=\tfrac{i}{2}[\gamma^\mu,\gamma^\nu]$ and $q$ is the momentum transferred to the photon. In the
limit $q^2\to0$ the form factors give the static moments, $F_1(0) = 1$, $F_2(0) = \atau$ and
$F_3(0) = 2m_\tau\dtau/e$. In the photon-induced process considered here both photons are quasi-real
($q^2\simeq 0$), and deviations from the SM are parametrised by constant couplings, $F_2 = \ka$ and $F_3 = \kta$, so
that
\begin{equation}
  \atau = \ka, \qquad \dtau = \frac{e\,\kta}{2m_\tau} = \kta \times \num{5.553e-15}\,\ecm .
  \label{eq:identification}
\end{equation} The couplings
describe the shift with respect to the SM at tree level. The SM value of $\atau$ which arises at loop level is not
included in the SM prediction used in this tudy.

\begin{figure}[tbp]
  \centering
  \includegraphics[width=0.85\textwidth]{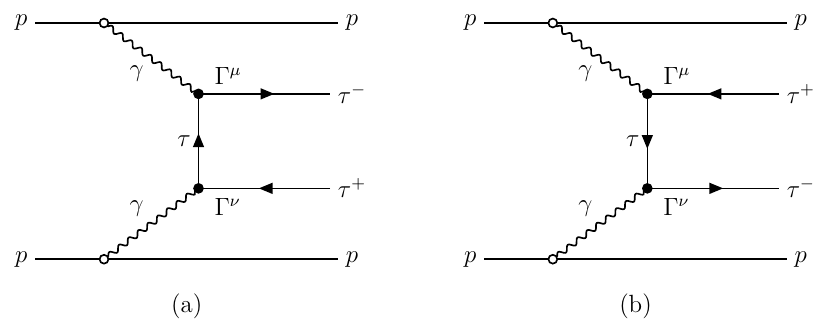}
  \caption{Leading-order diagrams for the elastic process $pp\to p\,\tau^+\tau^-p$: (a) $t$-channel and (b)
    $u$-channel $\tau$ exchange. The full circles denote the $\gamma\tau\tau$ vertex of Eq.~\eqref{eq:vertex}, which
    contains the anomalous couplings, and the open circles the elastic photon emission from the protons. In the
    proton-dissociative contributions one or both protons break up.}
  \label{fig:feynman}
\end{figure}

At leading order, $\gaga\to\tau^+\tau^-$ proceeds through $t$- and $u$-channel $\tau$ exchange
(Fig.~\ref{fig:feynman}), so that each diagram contains two $\gamma\tau\tau$ vertices of the form of
Eq.~\eqref{eq:vertex}, one for each incoming photon. The amplitude therefore receives contributions linear in the
couplings from each of the two vertices, and the cross section is a polynomial of fourth order in $(\ka,\kta)$.
Since $\kta$ is CP-odd, only even powers of $\kta$ contribute to CP-even quantities such as the total cross
section or the kinematic distributions used in this analysis, so that $\sigma(+\dtau)=\sigma(-\dtau)$.
The term linear in $\ka$ comes from the interference with the SM amplitude. It is suppressed by $m_\tau/\sqrt{\hat s}$
and leads to a small asymmetry between positive and negative values of $\atau$. Since the dipole terms in
Eq.~\eqref{eq:vertex} are proportional to the photon momentum, the anomalous terms grow with the invariant mass of
the $\tau$ pair, while the SM cross section falls.

%% file: q/sections/simulation.tex
\section{Event generation and detector simulation}\label{sec:simulation}

\subsection{Elastic signal}
The anomalous $\gamma\tau\tau$ vertex of Eq.~\eqref{eq:vertex} is implemented with \textsc{FeynRules}~2.3
\cite{Alloul:2013bka} and exported as a Universal FeynRules Output (UFO) model~\cite{Degrande:2011ua} where the
couplings $\ka$ and $\kta$ as free parameters. This model is merged with the TauDecay UFO
model~\cite{Hagiwara:2012vz} which describes the hadronic $\tau$ decays through effective vertices. The combined
model is inserted into \MG\ 3.5.12~\cite{Alwall:2014hca} with which the signal events are generated. The $\tau$
decays are thus included in the matrix element with full spin correlations. Both protons emit quasi-real photons and remain intact.
The photon-photon luminosity is taken from the \texttt{gamma}-UPC framework~\cite{Shao:2022cly,Crepet:2024dvv} implemented
in \MG, with
photon fluxes derived from the charge form factor of the proton (option \texttt{chff}). The probability that the
protons do not interact hadronically (soft survival) is included in impact-parameter space. Only hadronic decays are generated,
$\tau\to\nu_\tau+\{\pi,\ \pi\pi^0,\ \pi\pi^0\pi^0,\ 3\pi\}$, i.e.\ 16 combinations of decay channels with a
summed branching fraction of 54.6\% per $\tau$. All cross sections quoted below include these branching fractions.

In the TauDecay model~\cite{Hagiwara:2012vz} used in the merged UFO model, the hadronic decays are described by the
effective interaction
\begin{equation}
  \mathcal{L}_{\rm eff} = \frac{G_F}{\sqrt2}\,\cos\theta_c\;
  \bar\nu_\tau\gamma_\mu(1-\gamma^5)\tau\; J^\mu + {\rm h.c.},
  \label{eq:taudecay}
\end{equation}
where the hadronic current $J^\mu$ for the one- and three-prong channels considered here reads, up to an overall
phase,
\begin{align}
  \tau^-\to\pi^-\nu_\tau:&\quad J^\mu = f_\pi\, p_\pi^\mu ,\nonumber\\
  \tau^-\to\pi^-\pi^0\nu_\tau:&\quad J^\mu = \sqrt2\, B_\rho(Q^2)\,\big(p_{\pi^-}-p_{\pi^0}\big)^\mu ,
  \label{eq:currents}\\
  \tau^-\to(3\pi)^-\nu_\tau:&\quad J^\mu = \frac{4}{3f_\pi}\,B_{a_1}(Q^2)
  \Big(g^{\mu\nu}-\frac{Q^\mu Q^\nu}{Q^2}\Big)
  \Big[B_\rho(s_{13})\,(p_1-p_3)_\nu + B_\rho(s_{23})\,(p_2-p_3)_\nu\Big] .\nonumber
\end{align}
Here $Q$ is the total momentum of the hadronic system, $s_{ij}=(p_i+p_j)^2$, and $p_3$ denotes the pion whose charge
differs from that of the other two ($\pi^+$ in $\pi^-\pi^-\pi^+$, $\pi^-$ in $\pi^-\pi^0\pi^0$). The intermediate
vector and axial-vector resonances enter through normalised Breit-Wigner functions,
$B_\rho = [{\rm BW}_\rho + \beta\,{\rm BW}_{\rho'}]/(1+\beta)$ and $B_{a_1}={\rm BW}_{a_1}$ with
${\rm BW}_X(s) = m_X^2/[m_X^2 - s - i\sqrt{s}\,\Gamma_X(s)]$ and energy-dependent widths of the form
of Ref.~\cite{Kuhn:1990ad}. The parameter values implemented in the TauDecay model are $f_\pi=130.41$~MeV,
$m_\rho=775.5$~MeV, $\Gamma_\rho=149.1$~MeV, $m_{\rho'}=1465$~MeV, $\Gamma_{\rho'}=400$~MeV, $\beta=-0.145$,
$m_{a_1}=1230$~MeV and $\Gamma_{a_1}=420$~MeV. The one-prong modes $\pi\nu$, $\pi\pi^0\nu$ and $\pi\pi^0\pi^0\nu$ are
thus mediated by direct pion emission and by the $\rho$ and $a_1$ resonances, respectively, and the three-prong mode
$3\pi\nu$ by the $a_1\to\rho\pi$ chain. Since the full $\tau$ decay amplitudes are part of the matrix element, the
angular distributions of the decay products reflect the $\tau$ polarisations and their correlations, which are
modified by the anomalous couplings.

At generator level, both $\tau$ leptons are required to have
$p_T > 20$~GeV and $|y| < 2.5$, to be separated by $\Delta R > 0.4$, and the invariant mass of the pair has to
satisfy $\mtt > 100$~GeV. The multi-body decay chains require an improved phase-space integration\ifextrasec. Details and
validation are given in Appendix~\ref{app:validation}\fi. The generated events are passed through \textsc{Pythia}~8.316~\cite{Sjostrand:2014zea,Bierlich:2022pfr} which
decays the unstable hadrons (in the signal, the $\pi^0$ mesons from the $\tau$ decays) and provides the event record
for the detector simulation. Initial-state
radiation, multi-parton interactions and beam remnants are switched off, because the elastic photon-photon process
produces nothing besides the $\tau$ decay products in the central detector.

\subsection{Reweighting}
A single sample of \num{200000} events is generated at the anomalous base point $(\ka,\kta) = (0.01, 0.01)$ and
reweighted with \MG\ to 26 further points: $\ka \in \{0,\allowbreak \pm\num{2.5e-4},\allowbreak \pm\num{5e-4},\allowbreak \pm10^{-3},\allowbreak
\pm\num{2e-3},\allowbreak \pm\num{5e-3},\allowbreak \pm10^{-2}\}$ at $\kta=0$, $\kta\in\{\num{2.5e-4},\allowbreak\ldots,\allowbreak10^{-2}\}$ at $\ka=0$, and seven mixed points.
An anomalous base point populates the high-mass region much better than an SM base point. Since the same events are used for all points, the statistical fluctuations are strongly correlated between
them. The reweighting is closed at the level of \SI{0.3}{\percent}. The SM cross section obtained by reweighting,
0.03027~pb, agrees with a direct generation, 0.03017~pb (84~TeV).

\begin{figure}[tbp]
  \centering
  \includegraphics[width=0.7\textwidth]{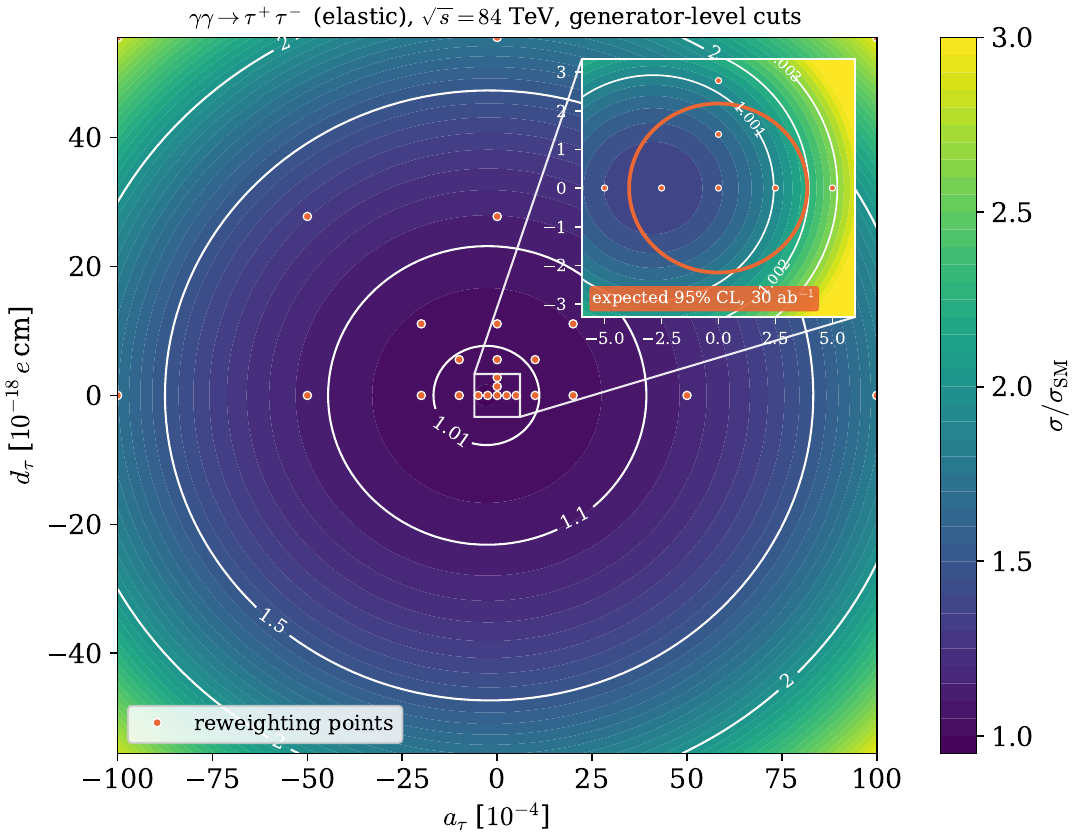}
  \caption{Elastic $\gaga\to\tau^+\tau^-$ cross section relative to the SM as a function of $\atau$ and
    $\dtau$ at $\sqrt{s} = 84$~TeV with the generator-level cuts, from the polynomial fit to the base point and the
    reweighted points (dots). The inset shows the central region with its own colour scale and the expected 95\% CL
    region for $\mathcal{L} = 30~\invab$.}
  \label{fig:xsmap}
\end{figure}

Figure~\ref{fig:xsmap} shows the generator-level cross section as a function of $\atau$ and $\dtau$, obtained from a
fourth-order polynomial fitted to the \res{xs-npts} simulated points, which it reproduces within \res{xs-fitdev}\%.
The cross section is symmetric in $\dtau$, as expected for a CP-even quantity, and the interference term shifts the
minimum slightly towards negative $\atau$. Close to the expected sensitivity the total cross section changes very
little: for $\atau = 5\times10^{-4}$ ($\dtau = 2.8\times10^{-18}\,\ecm$) it exceeds the SM value by only
\res{xs-r-a5}\% (\res{xs-r-d5}\%), and the expected 95\% CL region for $30~\invab$ (inset) lies within the contour
of a 0.2\% change.

\subsection{Proton-dissociative production}\label{sec:diss}
In addition to the elastic process, photons can be emitted inelastically, with one (single dissociation, sd) or both
(double dissociation, dd) protons breaking up. These contributions are not part of the \MG\ simulation and are
estimated with \SuperChic\ 5.4~\cite{Harland-Lang:2020veo}, which includes the soft survival factor for each
topology~\cite{Harland-Lang:2024zpn}. With the generator-level cuts of the signal, the cross sections relative to the
elastic one are $\sigma_{\rm sd}/\sigma_{\rm el}\approx 4.2$--4.7 and $\sigma_{\rm dd}/\sigma_{\rm el}\approx 0.6$ for the
four energies; the latter is strongly reduced by the small survival factor of the double-dissociative topology,
$\Stwo_{\rm dd}\approx 0.08$, compared with $\Stwo_{\rm sd}\approx 0.77$ and $\Stwo_{\rm el}\approx 0.95$.

The dissociated proton can produce additional tracks and spoil the exclusivity requirement. The efficiency of this
requirement for dissociative events is determined at generator level: unweighted \SuperChic\ events are showered with
\Pythia\ with the beam-remnant settings recommended by the \SuperChic\ authors, and the
probability that no additional charged particle is
reconstructed is computed with the tracking efficiency of the detector parametrisation, as for the DY background.
Elastic, single- and double-dissociative samples are processed identically, so that the efficiency ratios
$\varepsilon_{\rm diss}/\varepsilon_{\rm el}$ are largely independent of the $\tau$ decay modelling. In about half of
the single-dissociative events no additional track is reconstructed, since the proton remnant is emitted at very
forward rapidities or with low transverse momentum. The ratio $\varepsilon_{\rm sd}/\varepsilon_{\rm el}$ is
\res{epsratio-sd-low} for $\mtt < 1$~TeV and \res{epsratio-sd-high} above (\res{epsratio-dd-low} and
\res{epsratio-dd-high} for dd), as determined from dedicated samples at 84~TeV, which are used for all energies.

The dissociative contribution is included as a per-event factor depending on the generated $\mtt$,
$C(\mtt) = \sum_{\rm sd,dd}(\sigma_{\rm diss}/\sigma_{\rm el})(\mtt)\cdot(\varepsilon_{\rm diss}/\varepsilon_{\rm el})(\mtt)$,
separately for the SM part ($C_{\rm SM}$) and for the anomalous part ($C_{\rm NP}$, obtained from the cross sections at
$\atau = 0.01$). Since the cross-section ratios are evaluated consistently within \SuperChic, the dissociative
contribution is normalised to the elastic \MG\ prediction, which includes the \texttt{gamma}-UPC survival probability. The survival of the dissociative topologies enters only through their survival factors relative to the elastic one. 
In the signal region,  the dissociative contribution is about twice as large as the elastic one. For the four center-of- mass energies
and all $\mtt$ bins covered in this study, the dissociative contribution ranges between $\res{CSM-min}$ and $\res{CSM-max}$ for $C_{\rm SM}$ while  $\res{CNP-min}$ and $\res{CNP-max}$ for $C_{\rm NP}$.
The related values at 84~TeV center-of- mass energy are given in Table~\ref{tab:dissfactors} while those at the other center-of-mass energies are evaluated by using the efficiency ratios from 84~TeV and the cross-section ratios
at corresponding energies.

\begin{table}[tbp]
  \centering
  \small
  \caption{Proton-dissociative contribution at $\sqrt{s} = 84$~TeV in bins of the generated $\mtt$: cross-section ratios
    from \SuperChic\ (SM, including soft survival), exclusivity-efficiency ratios, and the resulting factors $C_{\rm SM}$ and
    $C_{\rm NP}$ for the SM and the anomalous parts (sum of sd and dd).}
  \input{q/tables/diss_factors}

  \label{tab:dissfactors}
\end{table}

\subsection{Backgrounds}
The main backgrounds are Drell--Yan $\tau$-pair production, $pp\to Z/\gamma^*\to\tau^+\tau^-$ (DY), and exclusive
$W$-pair production, $\gaga\to W^+W^-$ with $W\to\tau\nu_\tau$. Both are simulated at leading order with \MG. The DY
sample is generated with the NNPDF2.3 LO parton distributions~\cite{Ball:2012cx} and the same $\tau$-level cuts as the
signal ($\sigma = \res{xs-dy-84}$~pb at 84~TeV). The $\gaga\to W^+W^-$ sample uses the elastic \texttt{gamma}-UPC photon
fluxes, with the cuts applied to the $\tau$ leptons from the $W$ decays ($\sigma = \res{xs-ww-84}$~fb). In both
samples the $\tau$ leptons are decayed by \Pythia. For DY the parton shower, hadronisation and underlying event are
simulated with the default \Pythia\ settings, including initial-state radiation and multi-parton interactions. For
$\gaga\to W^+W^-$, the same settings as for the elastic signal are used. No DY event of the fully simulated sample passes the exclusivity requirement, so its
efficiency is factorised. The shape and all other selection efficiencies are taken from the \Delphes\ sample without the
exclusivity requirement, and the probability that no additional track is reconstructed, $\varepsilon_{\rm excl}$,
is computed from a large generator-level sample as a function of $p_T(\tau_1)+p_T(\tau_2)$, folding the charged
particles with the tracking efficiency of the detector parametrisation. After the full selection
$\varepsilon_{\rm excl}^{\rm DY} \approx \res{eps-excl-dy}$. Multijet production with jets misidentified as $\tauh$
has the same underlying-event activity as DY and an additional suppression from the misidentification rate squared therefore
it is neglected.

The detector response is simulated with \Delphes\ 3.5.0~\cite{deFavereau:2013fsa} using the FCC-hh
parametrisation, without pile-up. The tracking system is embedded in a 4~T solenoidal field and reconstructs
charged particles up to $|\eta| = 6$; no tracks are reconstructed below $p_T = 0.1$~GeV.
The exclusivity requirement relies on this wide coverage, and the same tracking efficiencies enter the
generator-level evaluation of $\varepsilon_{\rm excl}$ for DY and for the proton-dissociative contributions.

Two FCC-hh parametrisations are available in \Delphes: the \texttt{FCChh\_II} card (``nominal revisited'') and
the optimistic \texttt{FCChh\_I} card. Both cards share the tracker layout, the parametrisation of the track
momentum resolution, the reduction of the tracking efficiency below $p_T = 1$~GeV (factors between 0.99 and 0.93),
and the dual-readout calorimeter module. They differ in the tracking efficiency, the calorimeter resolutions and
the flavour-tagging efficiencies (Table~\ref{tab:cards}). For $p_T > 1$~GeV the tracking efficiency is 95\% for
$|\eta| < 2.5$, 90\% for $2.5 < |\eta| < 4$ and 86\% for $4 < |\eta| < 6$ in \texttt{FCChh\_II}, and 100\%, 95\% and
90\% in \texttt{FCChh\_I}. The parametrised $\tau$ tagging of \Delphes\ is not used here, and $b$-tagging enters
only through the $b$-jet veto, which removes a negligible fraction of the signal.

\begin{table}[tbp]
  \centering
  \small
  \caption{Differences between the two FCC-hh parametrisations of \Delphes\ ($E$ in GeV; both HCAL parametrisations
    include a noise term of 0.05~GeV). The tracker layout and the momentum resolution are identical; the muon
    tracking efficiency is also slightly higher in \texttt{FCChh\_I}.}
  \begin{tabular}{lll}
    \toprule
     & \texttt{FCChh\_II} (baseline) & \texttt{FCChh\_I} \\
    \midrule
    track efficiency, $p_T > 1$~GeV & 95\%, 90\%, 86\% & 100\%, 95\%, 90\% \\
    \quad ($|\eta| < 2.5$, $2.5$--4, 4--6) & & \\
    ECAL, $|\eta| < 4$ & $8.2\%/\sqrt{E} \oplus 0.15\% \oplus 0.31~{\rm GeV}/E$ & $3\%/\sqrt{E} \oplus 0.5\% \oplus 2~{\rm MeV}/E$ \\
    ECAL, $4 < |\eta| < 6$ & $30\%/\sqrt{E} \oplus 3.5\% \oplus 0.31~{\rm GeV}/E$ & as for $|\eta| < 4$ \\
    HCAL, $|\eta| < 4$ & $48\%/\sqrt{E} \oplus 1\%$ & $30\%/\sqrt{E} \oplus 1\%$ \\
    HCAL, $4 < |\eta| < 6$ & $100\%/\sqrt{E} \oplus 1\%$ & as for $|\eta| < 4$ \\
    $b$- and $\tau$-tagging efficiency & nominal & $\times 1.05$ \\
    \bottomrule
  \end{tabular}

  \label{tab:cards}
\end{table}

The analysis is performed for the four FCC-hh operating scenarios of Table~\ref{tab:scenarios}, with the
\texttt{FCChh\_II} card as the baseline. For each energy the full chain (signal with reweighting, proton-dissociative
factors, $\gaga\to WW$, and DY including $\varepsilon_{\rm excl}$) is repeated with identical selection and binning.
The efficiency ratios of the dissociative contributions are taken from the 84~TeV samples for each scenario. To assess the dependence on the detector performance, the chain is repeated at all four energies with the \texttt{FCChh\_I} card: the same generated
signal events are passed through \Delphes\ with this card, the $\gaga\to WW$ and DY samples are simulated again, and
$\varepsilon_{\rm excl}$ for DY and the dissociative efficiency ratios are recomputed with the tracking efficiencies of this card.
The resulting limits are discussed in Sec.~\ref{sec:systematics}.

%% file: q/tables/diss_factors.tex
\begin{tabular}{lccccc}
\toprule
$\mtt$ [GeV] & $\sigma_{\rm sd}/\sigma_{\rm el}$ & $\sigma_{\rm dd}/\sigma_{\rm el}$ & $\varepsilon_{\rm sd}/\varepsilon_{\rm el}$ & $C_{\rm SM}$ & $C_{\rm NP}$ \\
\midrule
100--300 & 4.30 & 0.60 & 0.41 & 1.94 & 2.31 \\
300--1000 & 4.07 & 0.62 & 0.41 & 1.85 & 1.82 \\
1000--3000 & 3.26 & 0.61 & 0.60 & 2.18 & 2.06 \\
3000--$\sqrt{s}$ & 2.74 & 0.90 & 0.60 & 1.96 & 2.43 \\
\bottomrule
\end{tabular}

%% file: q/sections/selection.tex
\section{Hadronic $\tau$ reconstruction and event selection}\label{sec:selection}
Hadronic $\tau$ decays are reconstructed with an algorithm modelled on the hadrons-plus-strips approach
\cite{CMS:2015pac,CMS:2018jrd}, applied directly to the particle-flow objects of \Delphes\ (\texttt{EFlowTrack} and
\texttt{EFlowPhoton}), independently of the jet clustering and of the parametrised $\tau$ tagging of \Delphes.
Tracks with $p_T > 5$~GeV serve as seeds. Photons are combined into $\pi^0$ candidates (``strips'') within
$|\Delta\eta| < 0.05$ and $|\Delta\phi| < 0.20$ for $E_T > 1$~GeV. The decay products are collected in a signal cone
of radius $R = 3~\mathrm{GeV}/p_T$, restricted to $0.05 \le R \le 0.20$, considering only tracks within $|\Delta z| < 1$~mm
of the leading track. Candidates are classified into the four decay modes: one prong, one prong plus $\pi^0$, three prongs
and three prongs plus $\pi^0$, and are required to have unit charge, a visible mass within a mode-dependent window
and an isolation requirement $\sum p_T < 2$~GeV of the charged particles within $\Delta R < 0.4$. All kinematic quantities are
computed in double precision, since the single-precision four-vectors stored by \Delphes\ lose accuracy for the
invariant masses of multi-TeV objects. In the signal sample the algorithm reconstructs \res{reco-eff}\% of the
hadronic $\tau$ decays within the acceptance, with the correct decay mode in \res{reco-mode}\% and the correct
number of prongs in \res{reco-prong}\% of the cases. Only \res{reco-fake-n} out of \res{reco-fake-d} candidates are
not matched to a generated $\tau$.

\begin{figure}[tbp]
  \centering
  \includegraphics[width=\textwidth]{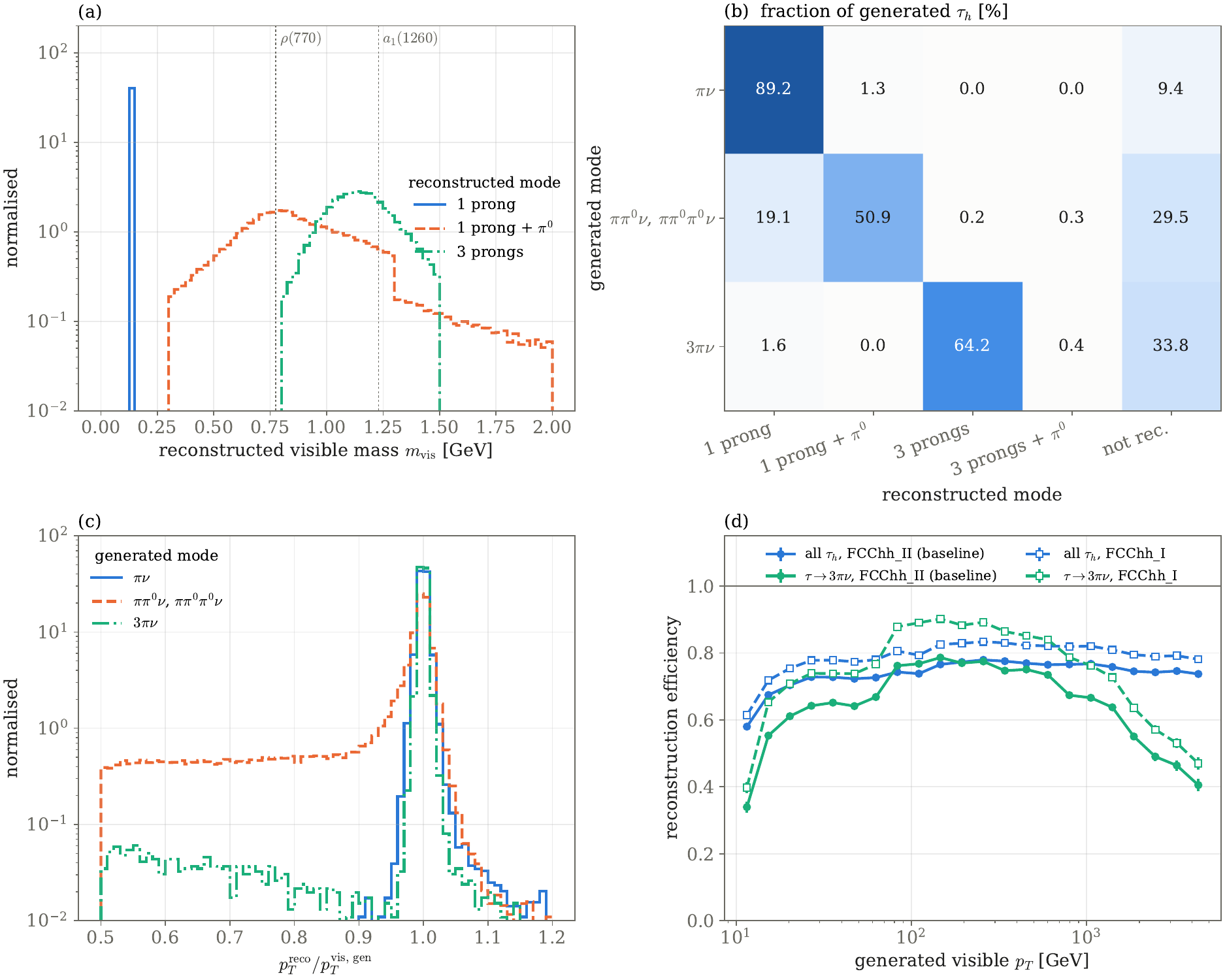}
  \caption{Performance of the hadronic $\tau$ reconstruction for the 84~TeV signal sample: (a) visible mass of the
    reconstructed candidates by decay mode, (b) fraction of generated $\tauh$ decays reconstructed in each mode,
    (c) ratio of the reconstructed to the generated visible $p_T$ by generated mode, and (d) reconstruction efficiency
    as a function of the generated visible $p_T$ for all $\tauh$ decays and for three-prong decays, for the
    \texttt{FCChh\_II} (baseline) and \texttt{FCChh\_I} detector parametrisations.}
  \label{fig:recoval}
\end{figure}

The performance of the reconstruction is illustrated in Fig.~\ref{fig:recoval} for the hadronic $\tau$ decays of the
84~TeV signal sample within $|\eta| < 2.5$ and visible $p_T > 10$~GeV, matched to the reconstructed candidates within
$\Delta R < 0.2$. In the visible-mass distributions (a), one-prong candidates without
$\pi^0$ sit at the pion mass, those with $\pi^0$ strips peak near the $\rho(770)$, and three-prong candidates are
spread around the $a_1(1260)$ mass. The sharp edges are the mode-dependent mass windows. The decay mode
(b) is reconstructed correctly for \res{rv-1p-ok}\% of the one-prong and \res{rv-3p-ok}\% of the three-prong decays.
Decays with $\pi^0$ mesons are reconstructed with the correct mode in \res{rv-pi0-ok}\% of the cases and as one-prong
candidates without $\pi^0$ in \res{rv-pi0-as1p}\%, namely when the photons fall below the strip threshold or are not
resolved. Migrations between one- and three-prong modes stay below \res{rv-3p-as1p}\%. The transverse-momentum response (c) peaks at unity, with a core
resolution of \res{rv-res-1p}\% (\res{rv-res-3p}\%) for one-prong (three-prong) decays, determined by the tracker
resolution. For decays with $\pi^0$ mesons the core resolution is \res{rv-res-pi0}\% and \res{rv-pi0-core}\% of the
candidates lie within $\pm10\%$ of the true visible $p_T$. The tail towards low values originates from photons that
are not included in the strips.

The reconstruction efficiency (d) is about \res{rv-eff-low}\% for visible $p_T$ between 30 and 100~GeV and
is still \res{rv-eff-high}\% above 1~TeV. The efficiency for three-prong decays decreases above about 1~TeV, from
\res{mw-3p-low}\% at 50--300~GeV to \res{mw-3p-high}\% above 2~TeV. The loss comes from the fixed three-prong mass
window which was designed to reject quark and gluon jets at moderate $p_T$. At multi-TeV transverse momenta the
tracker momentum resolution broadens the reconstructed visible mass beyond it. Without the window, the three-prong efficiency remains approximately constant up to the highest transverse
momenta. The baseline retains the mass windows, as required for the suppression of multijet backgrounds in a
realistic environment. Since the backgrounds considered here contain genuine $\tau$ leptons, the results are
conservative in this respect, and a transverse-momentum-dependent definition of the windows is left for future
optimisation. With the \texttt{FCChh\_I} card the efficiency is higher by about 5
percentage points for all $\tau_h$ and by up to 13 percentage points for three-prong decays over the full $p_T$ range, owing to the higher tracking efficiency.

\subsection{Event selection}
Events are required to contain at least two reconstructed $\tauh$ candidates. The leading candidate and the highest-$p_T$ candidate of opposite charge form the $\tau$ pair, and are required to have $p_T > 40$~GeV and 30~GeV, respectively. Events with an electron or muon with $p_T > 10$~GeV are rejected. The back-to-back topology of the photon-induced process
is exploited by requiring $\Delta\phi(\tau_1,\tau_2) > 2.7$. The exclusivity requirement demands that no
additional track with $p_T > 0.5$~GeV is found outside cones of $\Delta R = 0.4$ around the two $\tau$ candidates
and within $|\Delta z| < 1$~mm of either $\tau$ track, $N_{\rm trk}^{\rm extra} = 0$. Using the $z$ positions of both $\tau$ tracks,
rather than with an averaged vertex avoids a bias from the displaced $\tau$ decay vertices. This requirement
rejects processes with additional central activity, but it retains a large fraction of the dissociative signal
contributions, whose proton remnants are often emitted at very forward rapidities or with low transverse momentum. Finally, events containing a
b-tagged jet with $p_T > 30$~GeV are rejected. The selection efficiencies at 84~TeV are listed in
Table~\ref{tab:cutflow}. As can be seen from the table, \res{sr-eff-84}\% of the total generated base-sample signal events pass this selection. The selected events are characterised by the total transverse mass
\begin{equation}
  \begin{aligned}
    \mttot &= \sqrt{m_T^2(\tau_1,\tau_2) + m_T^2(\tau_1,E_T^{\rm miss}) + m_T^2(\tau_2,E_T^{\rm miss})}\,,\\
    m_T(i,j) &= \sqrt{2\,p_{T,i}\,p_{T,j}\,(1-\cos\Delta\phi_{ij})}\,,
  \end{aligned}
  \label{eq:mttot}
\end{equation}
built from the two $\tauh$ candidates and the missing transverse momentum. The statistical analysis uses the
$\mttot$ distribution above 150~GeV, which removes \res{y-sm-below150}\% of the SM
events, all of them at low mass where the anomalous couplings have a negligible effect.

The expected event yields for $30~\invab$ at 84~TeV are given in Table~\ref{tab:yields}. In the fit range
$\mttot > 150$~GeV, \res{y-smel-150} elastic and \res{y-smdiss-150} proton-dissociative SM events are expected, where the
dissociative contribution is \res{y-diss-ratio} times the elastic one after the exclusivity requirement. The \res{y-bkg-150} background  events correspond to \ \res{y-bkg-rel-el}\% of the elastic
and \res{y-bkg-rel-tot}\% of the total SM yield and are dominated by $\gaga\to WW$. The DY background is suppressed
by the exclusivity requirement to a negligible level. 

For couplings of the order of the expected sensitivity, $\atau = 5\times10^{-4}$ or $\dtau = 2.8\times10^{-18}\,\ecm$
(i.e.\ $\ka$ or $\kta = 5\times10^{-4}$), the anomalous contributions (elastic plus dissociative) increase the total yield by only \res{y-npa-rel-150}\%, less than the systematic uncertaint. The relative excess, however, is concentrated at high masses: above $\mttot = 1$~TeV they amount to
\res{y-npa-1tev} events on top of \res{y-smtot-1tev} SM events (\res{y-npa-rel-1tev}\%), and above 3~TeV they
exceed the SM expectation of \res{y-smtot-3tev} events by a factor of \res{y-npa-rel-3tev}, with a background of less
than one event. The sensitivity therefore originates almost entirely from the highest $\mttot$ bins where the
analysis is statistically limited and the backgrounds play no role.

\begin{table}[tbp]
  \centering
  \small
  \caption{Cumulative selection efficiencies at $\sqrt{s} = 84$~TeV with respect to the generated events (after the
    generator-level cuts), for the elastic signal and $\gaga\to WW$ processes and for DY process. The signal column refers to the base sample
    $(\ka,\kta) = (0.01,0.01)$, which is enriched in high-mass events; the last row adds the lower edge of the fit
    range. For DY the exclusivity requirement is applied through the factorised efficiency
    $\varepsilon^{\rm DY}_{\rm excl}$ (Sec.~\ref{sec:simulation}).}
  \input{q/tables/cutflow_84}

  \label{tab:cutflow}
\end{table}

\begin{table}[tbp]
  \centering
  \small
  \caption{Expected numbers of events in the signal region at $\sqrt{s} = 84$~TeV for $30~\invab$, in the fit range
    $\mttot > 150$~GeV and for $\mttot > 1$~TeV and $\mttot > 3$~TeV. The anomalous excess is the difference with respect to SM for the indicated
    coupling (the other coupling set to zero), including the elastic and the proton-dissociative contributions.}
  \input{q/tables/yields_84}

  \label{tab:yields}
\end{table}

%% file: q/tables/cutflow_84.tex
\begin{tabular}{lrrr}
\toprule
Selection & $\gaga\to\tau\tau$ & $\gaga\to WW$ & DY $\tau\tau$ \\
\midrule
generated & 200\,000 & 50\,000 & 100\,000 \\
$\geq 2$ $\tauh$ candidates (valid, isolated) & 50.33\% & 16.01\% & 11.03\% \\
opposite charge & 50.16\% & 15.86\% & 10.21\% \\
$p_T(\tau_1) > 40$, $p_T(\tau_2) > 30$~GeV & 35.64\% & 4.46\% & 2.62\% \\
lepton veto ($e/\mu$, $p_T > 10$~GeV) & 35.60\% & 4.46\% & 2.62\% \\
$\Delta\phi(\tau_1,\tau_2) > 2.7$ & 35.49\% & 2.79\% & 1.84\% \\
exclusivity: $N_{\rm trk}^{\rm extra} = 0$ & 35.43\% & 2.79\% & $\varepsilon^{\rm DY}_{\rm excl}$ (see text) \\
b-jet veto ($p_T > 30$~GeV) & 35.43\% & 2.79\% & -- \\
$\mttot > 150$~GeV (fit range) & 27.92\% & 1.81\% & -- \\
\bottomrule
\end{tabular}

%% file: q/tables/yields_84.tex
\begin{tabular}{lrrr}
\toprule
 & $\mttot > 150$~GeV & $\mttot > 1$~TeV & $\mttot > 3$~TeV \\
\midrule
SM $\gaga\to\tau\tau$, elastic & 41\,062 & 136 & 1.6 \\
SM $\gaga\to\tau\tau$, dissociative & 78\,092 & 291 & 3.0 \\
anomalous excess, $\atau = 5\times10^{-4}$ & 1\,399 & 122 & 10 \\
anomalous excess, $\dtau = 2.8\times10^{-18}\,\ecm$ & 1\,266 & 120 & 10 \\
$\gaga\to WW$ & 951 & 8.4 & $<0.1$ \\
DY $\tau\tau$ & 80 & 0.1 & $<0.1$ \\
\bottomrule
\end{tabular}

%% file: q/sections/statistics.tex
\section{Statistical analysis}\label{sec:statistics}

\begin{figure}[tbp]
  \centering
  \includegraphics[width=0.6\textwidth]{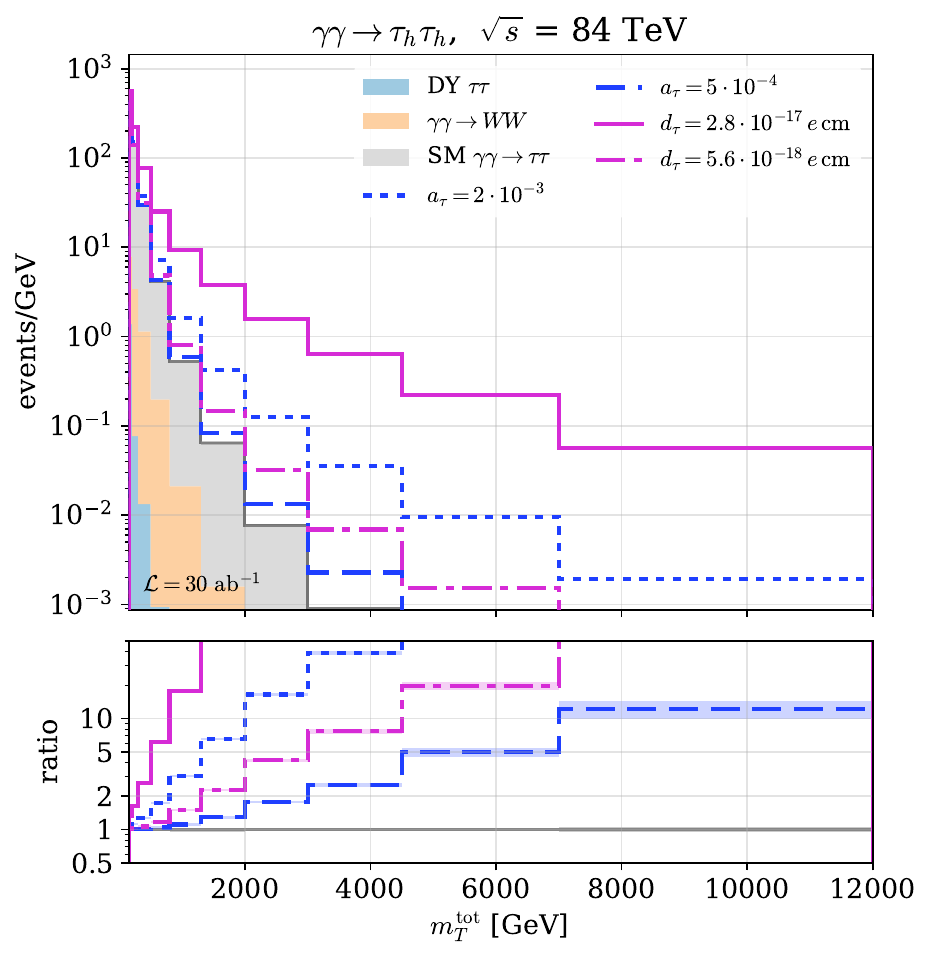}
  \caption{Distribution of $\mttot$ after the full selection at $\sqrt{s} = 84$~TeV for $\mathcal{L}=30~\invab$:
    elastic SM $\gaga\to\tau\tau$ expectation and the backgrounds (stacked) together with the elastic prediction for representative values of
    $\atau$ and $\dtau$; the proton-dissociative contribution is not shown. The lower panel shows
    the ratio to the SM expectation.}
  \label{fig:mttot}
\end{figure}

The discriminating variable is the total transverse mass $\mttot$ of Eq.~\eqref{eq:mttot} which serves as a
proxy for $\mtt$. Its distribution is shown in
Fig.~\ref{fig:mttot} for the SM and for representative values of $\atau$ and $\dtau$. The anomalous contributions
dominate for $\mttot \gtrsim 1$~TeV where the SM expectation is small.

The $\mttot$ distribution after the selection is binned with edges at 150, 200, 300, 500, 800, 1300, 2000, 3000,
4500, 7000 and 12000~GeV where the last bin includes the overflow. The binning was optimised for the expected
sensitivity with the requirement of sufficient Monte Carlo (MC) statistics: every bin contains at least 700
events of the simulated base sample (generated at $(\ka,\kta) = (0.01, 0.01)$ and reweighted to the SM and to the
other coupling points), and the relative MC uncertainty of the SM prediction is at most 7\% (reached in the
highest-mass bin at 72~TeV) and is included in the test statistic. In each bin $k$ the expected number of events is described by a polynomial in the couplings,
\begin{equation}
  N_k(\ka,\kta) = \sum_{i+j\le 4,\ j\ {\rm even}} c^{(k)}_{ij}\,\ka^i\,\kta^j ,
\end{equation}
whose nine coefficients are determined from the 27 simulated points (base point and 26 reweighted points) by a fit
weighted with the MC uncertainties. The polynomial reproduces the simulated points within their uncertainties.

Expected limits are derived with Asimov data~\cite{Cowan:2010js} $n_k$ corresponding to the SM prediction plus
backgrounds. Since the
most sensitive high-mass bins have SM expectations $n_k \ll 1$, the Poisson likelihood-ratio test
statistic~\cite{Baker:1983tu} is used, with the variance
inflated to account for a relative systematic uncertainty $\delta$, uncorrelated between bins, and for the MC
uncertainty $\sigma_{{\rm MC},k}$ of the SM prediction:
\begin{equation}
  q(\ka,\kta) = \sum_k 2\left[N_k - n_k + n_k\ln\frac{n_k}{N_k}\right]
                \frac{n_k}{n_k + (\delta\, n_k)^2 + \sigma_{{\rm MC},k}^2} .
  \label{eq:q}
\end{equation}
The factor multiplying the Poisson term is an approximation: in the Gaussian limit it is equivalent to profiling an
uncorrelated nuisance parameter in each bin, and it avoids a likelihood fit with one nuisance parameter per bin.
The signal prediction includes the elastic and the proton-dissociative contributions,
$N_k = N_k^{\rm el} + N_k^{\rm diss}$. The modelling uncertainty of the dissociative part is described by two nuisance
parameters, a normalisation $\theta_n$ and a slope $\theta_s$ in $\ln\mtt$,
$N_k^{\rm diss} \to N_k^{\rm diss}\,(1 + \theta_n + \theta_s\, t(\mtt))$ with $t$ varying linearly in $\ln\mtt$ from $-1$
at 100~GeV to $+1$ at 10~TeV. Both are constrained by Gaussian terms with widths of 25\%, which are added to
Eq.~\eqref{eq:q}, and are profiled for each value of the couplings. The slope parameter prevents the normalisation,
which is effectively calibrated by the high-statistics low-mass bins, from being extrapolated without uncertainty to the
high-mass region that drives the sensitivity.

For large $n_k$ Eq.~\eqref{eq:q} reduces to the usual Gaussian $\chi^2$. For the high-mass bins the Gaussian
approximation would overestimate the sensitivity by 7--8\%. A systematic uncertainty of $\delta = 5\%$ is assumed.
One-dimensional 95\% CL intervals are obtained from $q < 3.84$ with the other coupling set to zero, and
two-dimensional regions from $q < 5.99$, assuming the asymptotic $\chi^2$ distribution of $q$~\cite{Cowan:2010js}.

%% file: q/sections/results.tex
\section{Results}\label{sec:results}

\begin{figure}[tbp]
  \centering
  \includegraphics[width=\textwidth]{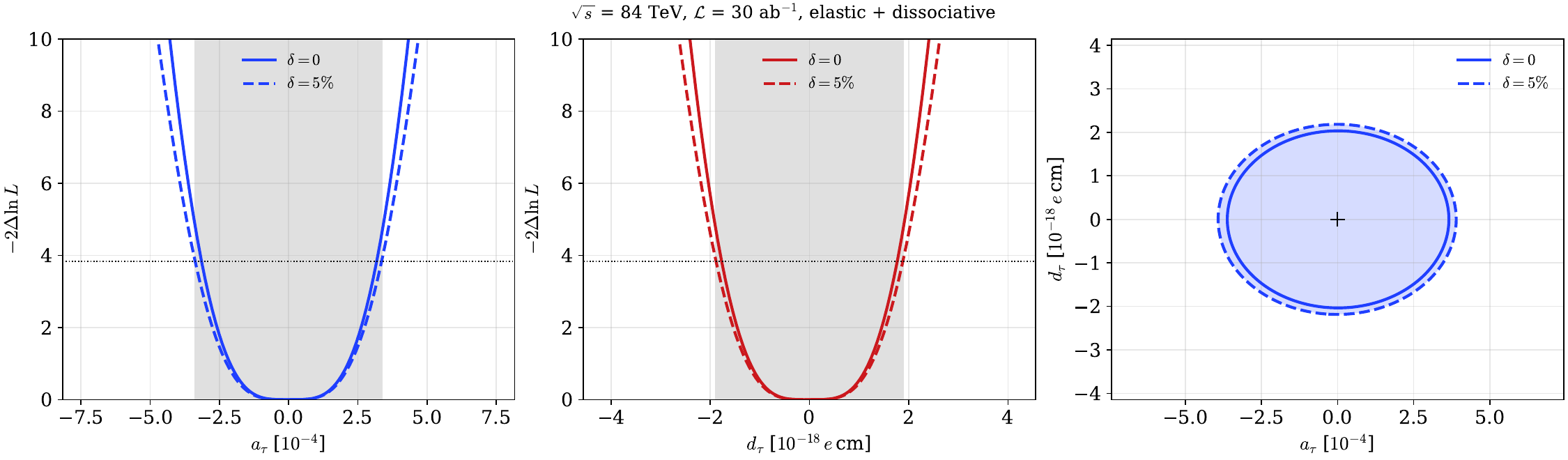}
  \caption{Expected $-2\Delta\ln L$ as a function of $\atau$ (left) and $\dtau$ (middle), and the 95\% CL region in
    the $(\atau,\dtau)$ plane (right), at $\sqrt{s}=84$~TeV for $\mathcal{L}=30~\invab$, including the
    proton-dissociative contribution, for a systematic uncertainty per bin of $\delta = 5\%$ (dashed, baseline) and
    $\delta = 0$ (solid). The dotted line marks $-2\Delta\ln L = 3.84$ and the shaded bands the baseline 95\% CL
    intervals.}
  \label{fig:limits84}
\end{figure}

For the baseline scenario at $\sqrt{s} = 84$~TeV and an integrated luminosity of $30~\invab$, the expected 95\% CL
intervals, obtained for one coupling at a time, are
\begin{equation}
  \res{ataulo-84-30}\times10^{-4} < \atau < +\res{atauhi-84-30}\times10^{-4}, \qquad
  |\dtau| < \res{dtau-84-30}\times10^{-18}\,\ecm .
\end{equation}
The corresponding likelihood scans and the two-dimensional region are shown in Fig.~\ref{fig:limits84}. The interval in $\atau$ is nearly symmetric: the interference term, which is suppressed by
$m_\tau/\sqrt{\hat s}$, is small in the high-mass bins that drive the sensitivity, and the remaining asymmetry of a
few per cent is within the MC statistical precision of the linear term. These intervals include the
proton-dissociative contribution, the backgrounds, the MC statistical uncertainty and a systematic uncertainty of
$\delta = 5\%$ per bin. Figure~\ref{fig:limits84} also shows the result without the per-bin systematic uncertainty
($\delta = 0$; the MC statistical uncertainty and the uncertainties of the dissociative contribution are kept),
where the intervals shrink to
$\res{s0-alo-84-30}\times10^{-4} < \atau < +\res{s0-ahi-84-30}\times10^{-4}$ and
$|\dtau| < \res{s0-d-84-30}\times10^{-18}\,\ecm$, i.e.\ by \res{s0-dgain-84-30}\%. The systematic uncertainty thus has a minor effect, because it affects
mainly the low-mass bins with large event counts, whereas the sensitivity comes from the high-mass bins, which are
statistically limited. Its effect at all energies and integrated luminosities is discussed below.
Including the
dissociative events as signal improves the limits by \res{diss-gain-min}--\res{diss-gain-max}\% with respect to the
elastic process alone.
These projections improve on the current best bounds from CMS~\cite{CMS:2024skm} by factors of about
\res{cms-a-ratio-84-30} for $\atau$ and \res{cms-d-ratio-84-30} for $\dtau$. The projected $\atau$ sensitivity is
smaller than the SM value itself, $\atau^{\rm SM}\simeq 1.2\times10^{-3}$ \cite{Eidelman:2007sb}.

\subsection{FCC-hh energy scenarios}
The expected limits for the four operating scenarios of FCC-hh suggested in Ref.~\cite{Zimmermann:2024fcchh} are summarised in
Table~\ref{tab:limits}.  Both the SM cross section and the signal-region efficiency increase with $\sqrt{s}$. The cross
section increases because a given $\tau$-pair mass requires smaller photon momentum fractions,
$x \approx \mtt/\sqrt{s}$, where the photon flux is larger. The efficiency evaluated for the base sample increases
mainly through the transverse-momentum requirements ($p_T > 40$ and 30~GeV, compared with 20~GeV at generator level)
which are passed by a larger fraction of the harder $\tau$ leptons at higher energies. All other requirements have
nearly energy-independent efficiencies.

\begin{figure}[tbp]
  \centering
  \includegraphics[width=0.6\textwidth]{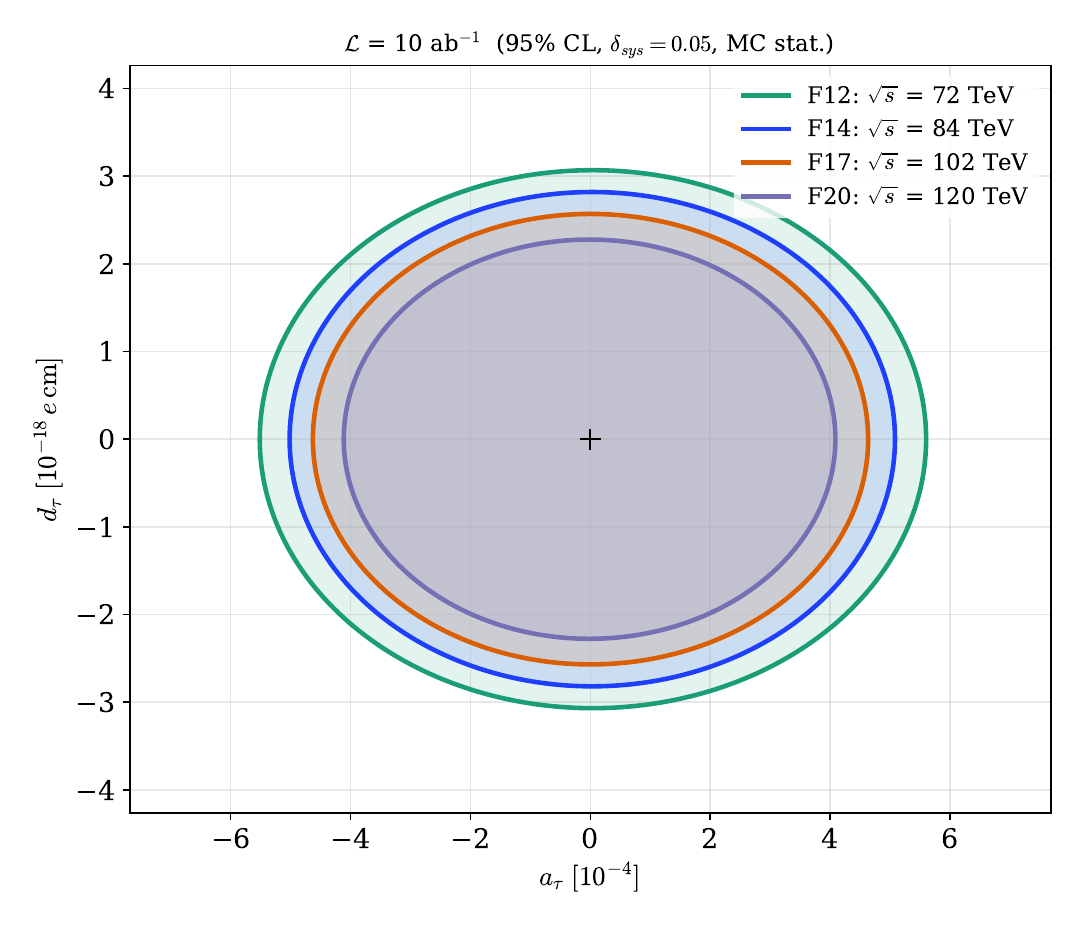}
  \caption{Expected 95\% CL regions ($-2\Delta\ln L < 5.99$) in the $(\atau,\dtau)$ plane for the four FCC-hh
    scenarios of Table~\ref{tab:scenarios} at a common integrated luminosity of $\mathcal{L}=10~\invab$, including
    the proton-dissociative contribution, the backgrounds, a systematic uncertainty of $\delta = 5\%$ per bin and the
    MC statistical uncertainty. The cross marks the SM point.}
  \label{fig:contours}
\end{figure}

Figure~\ref{fig:contours} compares the two-dimensional regions at a common integrated luminosity of $10~\invab$. The
regions shrink with increasing collider energy and are nearly elliptical and they extend to
$|\atau| \approx \res{kappa-a-84-10}\times10^{-4}$ and $|\dtau| \approx \res{dtau-84-10}\times10^{-18}\,\ecm$ at 84~TeV.
Converted with Eq.~\eqref{eq:identification}, the limit on $\dtau$ corresponds to
$|\kta| \lesssim \res{kappa-t-84-10}\times10^{-4}$, nearly the same value as for $|\ka| = |\atau|$ when expressed in terms of
the couplings, the regions are nearly circular. Since the observables used here are CP-even and the anomalous
contribution is dominated by the term that is quadratic in the couplings, the magnetic and the electric dipole couplings
produce nearly identical distortions of the $\mttot$ spectrum; the analysis therefore constrains approximately
$\ka^2 + \kta^2 = \atau^2 + (2m_\tau\dtau/e)^2$ and cannot distinguish between the two moments.

\begin{figure}[tbp]
  \centering
  \includegraphics[width=\textwidth]{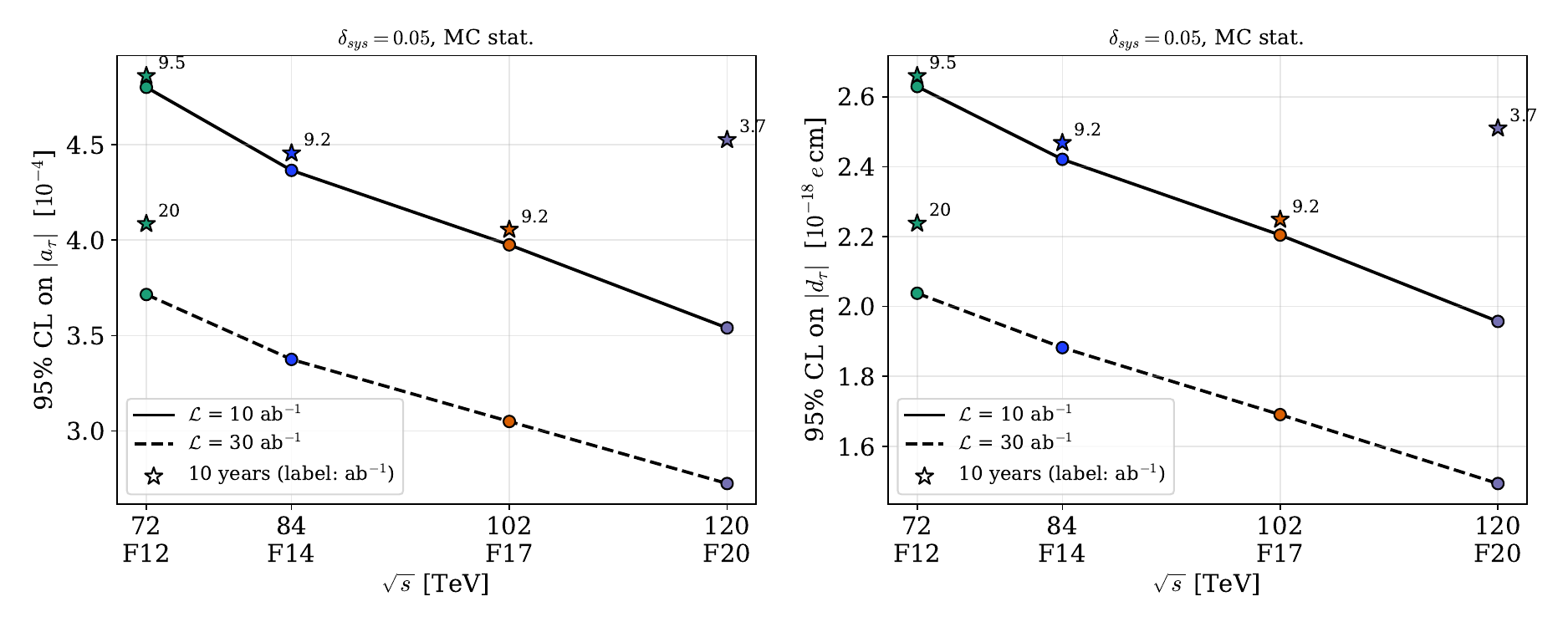}
  \caption{Expected 95\% CL limits on $|\atau|$ (left) and $|\dtau|$ (right) as a function of $\sqrt{s}$ at fixed
    luminosities of 10 and $30~\invab$ (lines) and for ten years of operation at the nominal luminosity of each scenario (stars,
    labelled with the integrated luminosity in $\invab$).}
  \label{fig:vssqrts}
\end{figure}

The dependence on the collider energy is shown in Fig.~\ref{fig:vssqrts}. At a fixed integrated luminosity the limits
improve monotonically with $\sqrt{s}$, by \res{gain-step-min}--\res{gain-step-max}\% from one scenario to the next.
Between 72 and 120~TeV the limit on $|\dtau|$ improves by \res{gain-1}\% at $1~\invab$, by \res{gain-10}\% at
$10~\invab$ and by \res{gain-30}\% at $30~\invab$. The gain grows with the luminosity because the additional
sensitivity comes from the highest masses where the event counts are smallest. At
$10~\invab$ the 72~TeV scenario would need about \res{lumi-equiv} times as much integrated luminosity to reach the
sensitivity of the 120~TeV scenario.

\begin{figure}[tbp]
  \centering
  \includegraphics[width=\textwidth]{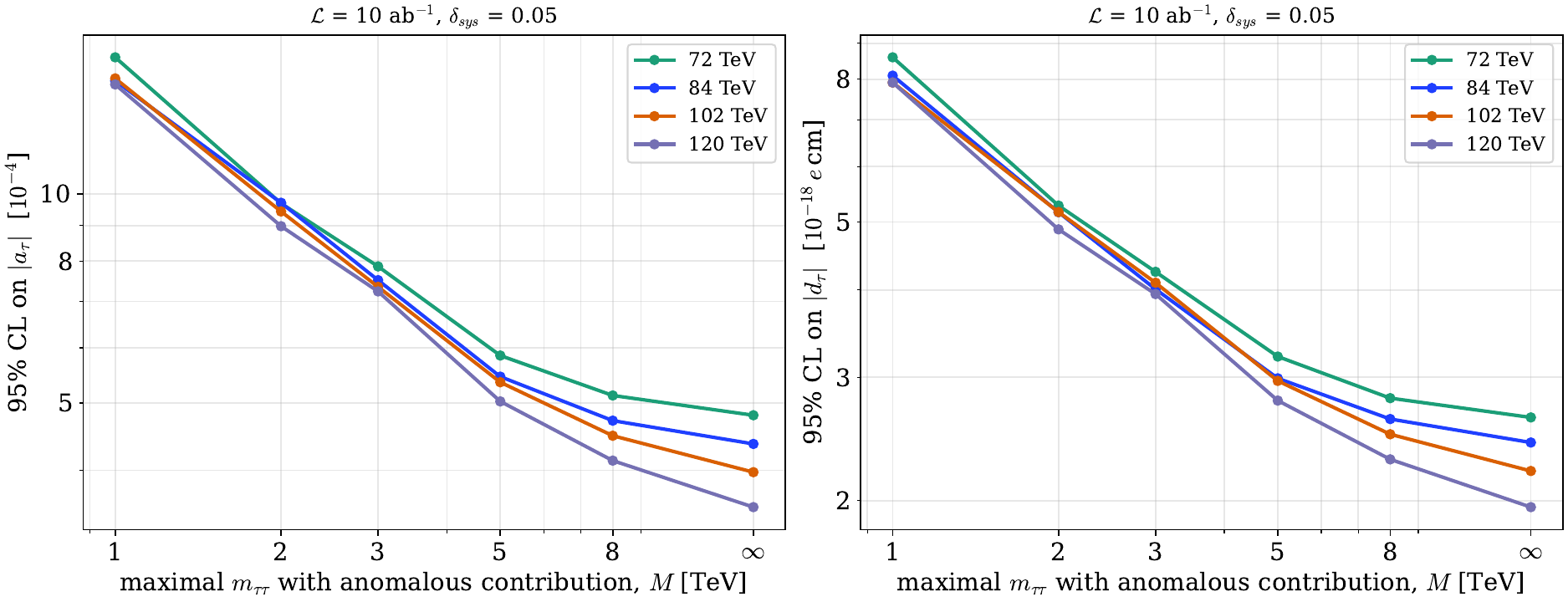}
  \caption{Expected 95\% CL limits on $|\atau|$ (left) and $|\dtau|$ (right) at $\mathcal{L} = 10~\invab$
    for the four scenarios when the anomalous contribution is restricted to $\mtt < M$ (events with larger generated $\mtt$ are
    treated as SM). The rightmost points ($M = \infty$) correspond to the unrestricted result.}
  \label{fig:mttmax}
\end{figure}

To identify the mass region that drives the sensitivity, the anomalous contribution is restricted to $\tau$-pair masses
below an upper cut-off $M$. The cut-off is applied to the generated invariant mass $\mtt$. Events with $\mtt > M$
are kept, but their weights are set to the SM value for all coupling points, so that they contribute only as SM
background. Figure~\ref{fig:mttmax} shows the resulting limits at $\mathcal{L} = 10~\invab$ as a function of $M$.
For $M = 3$~TeV, the limits on $|\dtau|$ become weaker by \res{clip3-worsen-min}--\res{clip3-worsen-max}\%. For
$M = 1$~TeV, they become weaker by a factor of \res{clip1-factor-min}--\res{clip1-factor-max}. The gain from a
higher collider energy also comes almost entirely from the multi-TeV region. Between $\sqrt{s} = 72$ and 120~TeV,
the limit on $|\dtau|$ improves by \res{gain-10}\% without restriction, but only by \res{clip2-gain}\% for
$M = 2$~TeV. The cut-off $M$ can also be interpreted as the validity range of the effective description. If the new
physics responsible for the anomalous couplings has a mass scale of a few TeV or below, the limits for the
corresponding value of $M$ should be used.

\begin{figure}[tbp]
  \centering
  \includegraphics[width=\textwidth]{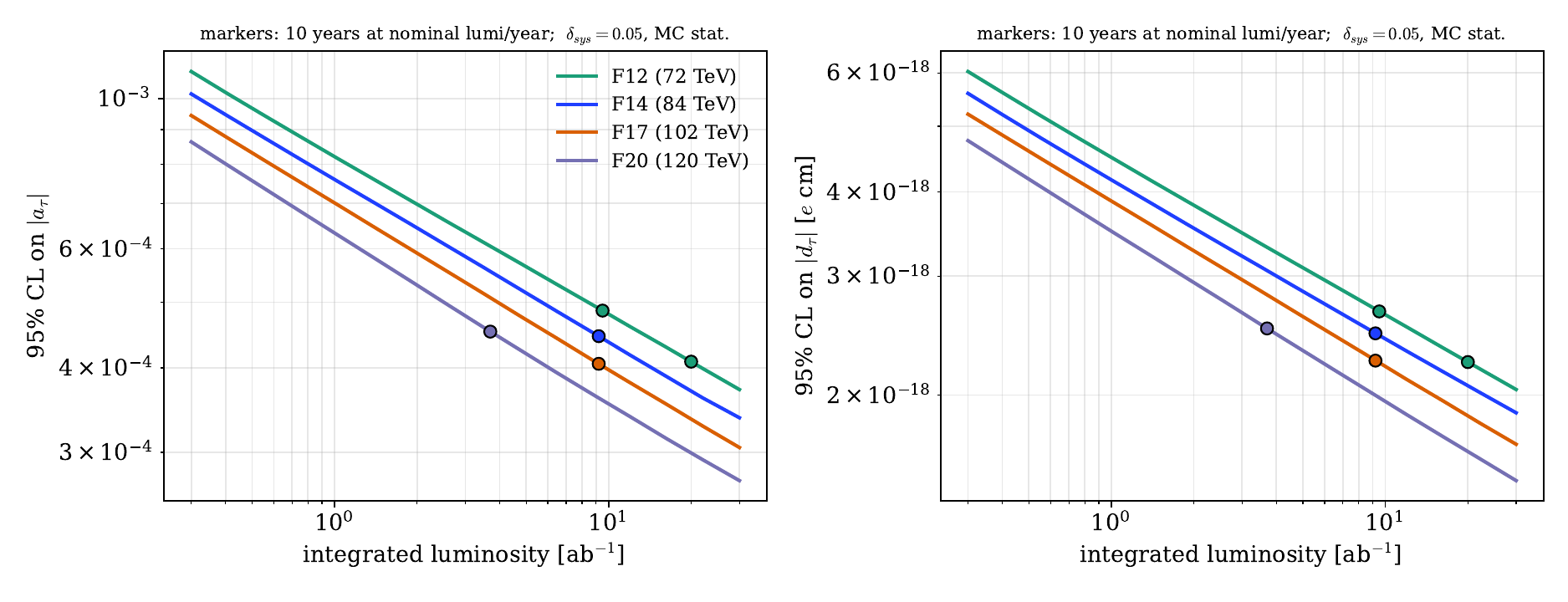}
  \caption{Expected 95\% CL limits on $|\atau|$ (left) and $|\dtau|$ (right) as a function of the integrated
    luminosity; markers indicate ten years of operation at the nominal annual luminosity of each scenario.}
  \label{fig:vslumi}
\end{figure}

Figure~\ref{fig:vslumi} shows the limits as a function of the integrated luminosity. For all energies they scale
approximately as $\mathcal{L}^{-1/4}$, with fitted exponents between \res{lscale-min} and \res{lscale-max}, as
expected for a statistically limited measurement of a contribution that is quadratic in the couplings. The slightly
flatter dependence at the highest luminosities reflects the growing weight of the systematic uncertainty and of the
uncertainties of the dissociative contribution. Because of this weak dependence, the differences in the annual
luminosity of the scenarios are as important as  the difference in their energies. For ten years of operation at the nominal annual
luminosity, the expected limits range from $|\atau| < \res{ten-a-best}\times10^{-4}$ and
$|\dtau| < \res{ten-d-best}\times10^{-18}\,\ecm$, obtained for the high-current option of the 72~TeV scenario with
$20~\invab$, to $|\atau| < \res{ten-a-worst}\times10^{-4}$ and $|\dtau| < \res{ten-d-worst}\times10^{-18}\,\ecm$ for
its low-current option. The 102~TeV scenario reaches a sensitivity similar to that of the high-current 72~TeV option
with less than half of its integrated luminosity, while the 120~TeV scenario, with $3.7~\invab$ in ten years, does
not benefit from its higher energy.

\begin{table}[tbp]
  \centering
  \small
  \setlength{\tabcolsep}{4pt}
  \caption{Expected 95\% CL limits for the FCC-hh scenarios (one coupling at a time, $\delta = 5\%$), including the
    proton-dissociative contribution as signal; the last column gives the $|\dtau|$  limit for the elastic process alone.
    $\sigma_{\rm SM}$ is the elastic SM cross section of $\gaga\to\tau^+\tau^-$ with both $\tau$
    leptons decaying hadronically, after the generator-level cuts of Sec.~\ref{sec:simulation}, and the signal-region (SR) efficiency is given for the base sample after all
    selection requirements except the fit range cut.}
  \input{q/tables/limits}

  \label{tab:limits}
\end{table}

%% file: q/tables/limits.tex
\begin{tabular}{lccllcc}
\toprule
Scenario & $\sigma_{\rm SM}$ [fb] & SR eff. [\%] & $\mathcal{L}$ [ab$^{-1}$] & $\atau$ [$10^{-4}$] & $|\dtau|$ [$10^{-18}\,\ecm$] & $|\dtau|$, el.\ only \\
\midrule
F12 (72 TeV) & 28.4 & 33.8 & 10 & $[{-}4.72,\ {+}4.80]$ & 2.63 & 2.90 \\
 &  &  & 30 & $[{-}3.67,\ {+}3.71]$ & 2.04 & 2.23 \\
 &  &  & 9.5 (10 yr) & $[{-}4.78,\ {+}4.86]$ & 2.66 & 2.94 \\
 &  &  & 20 (10 yr) & $[{-}4.03,\ {+}4.08]$ & 2.24 & 2.45 \\
\midrule
F14 (84 TeV) & 30.3 & 35.4 & 10 & $[{-}4.31,\ {+}4.36]$ & 2.42 & 2.68 \\
 &  &  & 30 & $[{-}3.38,\ {+}3.36]$ & 1.88 & 2.06 \\
 &  &  & 9.2 (10 yr) & $[{-}4.39,\ {+}4.45]$ & 2.47 & 2.73 \\
\midrule
F17 (102 TeV) & 32.9 & 37.3 & 10 & $[{-}3.98,\ {+}3.97]$ & 2.20 & 2.48 \\
 &  &  & 30 & $[{-}3.05,\ {+}3.04]$ & 1.69 & 1.89 \\
 &  &  & 9.2 (10 yr) & $[{-}4.06,\ {+}4.06]$ & 2.25 & 2.53 \\
\midrule
F20 (120 TeV) & 35.6 & 39.0 & 10 & $[{-}3.54,\ {+}3.51]$ & 1.96 & 2.26 \\
 &  &  & 30 & $[{-}2.73,\ {+}2.67]$ & 1.49 & 1.71 \\
 &  &  & 3.7 (10 yr) & $[{-}4.52,\ {+}4.52]$ & 2.51 & 2.93 \\
\bottomrule
\end{tabular}

%% file: q/sections/systematics.tex
\label{sec:systematics}

\begin{figure}[tbp]
  \centering
  \includegraphics[width=\textwidth]{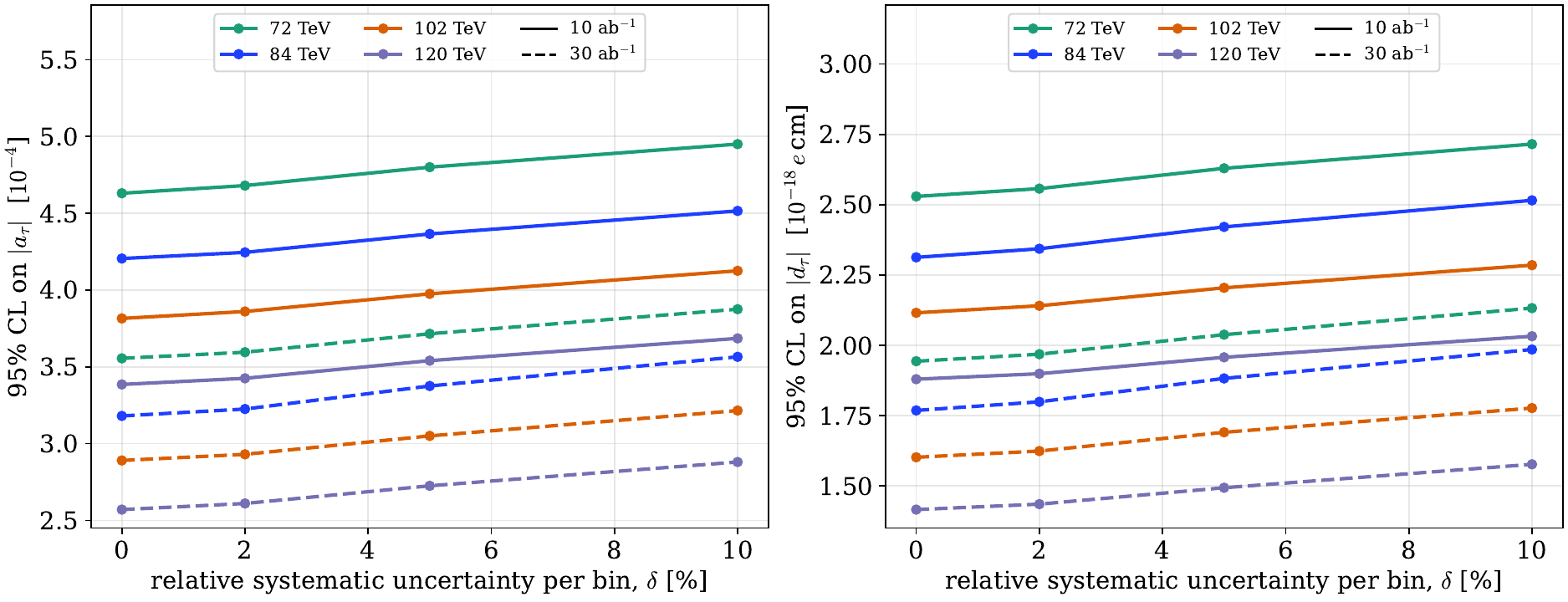}
  \caption{Expected 95\% CL limits on $|\atau|$ (left) and $|\dtau|$ (right) as a function of the relative systematic
    uncertainty per bin, for $10~\invab$ (solid lines) and $30~\invab$ (dashed).}
  \label{fig:syst}
\end{figure}

The baseline assumes a relative systematic uncertainty of $\delta = 5\%$ per bin, uncorrelated between bins. Because
the high-mass bins are statistically limited, the limits depend only weakly on this assumption (Table~\ref{tab:syst}
and Fig.~\ref{fig:syst}). At 84~TeV and $10~\invab$ the limit on $|\dtau|$ varies between \res{dtau-84-10} for
$\delta = 5\%$ and the values for $\delta = 0$ and 10\% given in the table, i.e.\ by less than 10\% over the full
range.

\begin{table}[tbp]
  \centering
  \small
  \caption{Expected 95\% CL limits on $|\dtau|$ ( and on $|\atau|$ in paranthesis)  for different values of the relative systematic uncertainty $\delta$ per bin (one coupling at a time).}
  \input{q/tables/syst}

  \label{tab:syst}
\end{table}

The dissociative events are included as signal with a 25\% normalisation and a 25\% slope uncertainty. Table~\ref{tab:disslimits} compares this baseline with the elastic process alone, with a
doubled slope uncertainty and with a conservative treatment in which the dissociative events are treated as SM-only
background without sensitivity to the couplings. Including them as signal improves the limits by
\res{diss-gain-min}--\res{diss-gain-max}\% while treating them as background degrades them by
\res{dissB-loss-min}--\res{dissB-loss-max}\% relative to the elastic process alone. The exclusivity efficiency of
the dissociative events relies on the \Pythia\ modelling of the proton remnant and is evaluated at generator level.

\begin{table}[tbp]
  \centering
  \small
  \caption{Expected 95\% CL limits on $|\dtau|$ at $10~\invab$ for different treatments of the proton-dissociative
    contribution.}
  \input{q/tables/diss_limits}

  \label{tab:disslimits}
\end{table}

With the optimistic \texttt{FCChh\_I} card  the signal is simulated from the same generated
and showered events, so that the differences are entirely due to the detector response. At 84~TeV the signal-region efficiency of the signal increases from \res{cardII-sr-84}\% to \res{cardI-sr-84}\%. The whole gain comes from the reconstruction of the two $\tauh$ candidates where the $\tauh$ reconstruction efficiency
rises from \res{cardII-reco-eff}\% to \res{cardI-reco-eff}\%, most strongly for three-prong decays
(\res{cardII-reco-3p}\% $\to$ \res{cardI-reco-3p}\%), for which all three tracks have to be reconstructed, and for
one-prong decays from \res{cardII-reco-1p}\% to \res{cardI-reco-1p}\%, while the relative efficiencies of all
subsequent requirements are unchanged. The exclusivity requirement hardly depends on the detector card:
the efficiency ratios $\varepsilon_{\rm diss}/\varepsilon_{\rm el}$ change by about \res{cardI-diss-eps-maxdev}\%,
and $\varepsilon_{\rm excl}^{\rm DY}$ agrees within its statistical uncertainty. The expected limits improve from $|\dtau| < \res{cardII-d-84-10}$ (\res{cardII-d-84-30}) to \res{cardI-d-84-10}
(\res{cardI-d-84-30})$\times10^{-18}\,\ecm$ at 10 (30)~$\invab$, i.e.\ by \res{cardI-gain-84-10}
(\res{cardI-gain-84-30})\%, with $\res{cardI-alo-84-10}\times10^{-4} < \atau < \res{cardI-ahi-84-10}\times10^{-4}$
at $10~\invab$. The small effect follows from the quadratic coupling dependence of the anomalous contribution in the high-$\mttot$
bins, which makes the limits scale roughly with the inverse fourth root of the efficiency ($1.15^{-1/4}\simeq0.97$). At the other energies the improvement is
\res{cardI-gain-min}--\res{cardI-gain-max}\% for all integrated luminosities considered, for a relative increase of
the signal-region efficiency of \res{cardI-sr-rel-min}--\res{cardI-sr-rel-max}\%. The baseline \texttt{FCChh\_II}
results are therefore conservative with respect to the detector performance.

The simulation does not include pile-up. With $\mathcal{O}(1000)$ simultaneous interactions expected at the FCC-hh, a
veto on additional tracks from the primary vertex requires a very precise vertex reconstruction, including timing
information, and/or the tagging of the intact protons in forward spectrometers, as done at the LHC with the CMS-TOTEM
Precision Proton Spectrometer and the ATLAS Forward Proton detectors~\cite{CMS:2014sdw,Adamczyk:2015cjy} for
photon-induced lepton-pair production~\cite{CMS:2018uvs,ATLAS:2020mve}. Such detectors are not yet part of the FCC-hh
reference detector design. Proton tagging would select the elastic process and remove the dissociative contribution which is
included as signal in the baseline; the limits would than correspond to the elastic-only values in Table~\ref{tab:disslimits} . The exclusivity efficiencies used here
should therefore be regarded as optimistic.
A full simulation of pile-up, including its effect on the $\tau$ isolation, is beyond the scope of this paper.

%% file: q/tables/syst.tex
\begin{tabular}{llcccc}
\toprule
 & & \multicolumn{4}{c}{$|\dtau|$ [$10^{-18}\,\ecm$]  ($|\atau|$ [$10^{-4}$])} \\
$\sqrt{s}$ & $\mathcal{L}$ [ab$^{-1}$] & $\delta = 0\%$ & $\delta = 2\%$ & $\delta = 5\%$ & $\delta = 10\%$ \\
\midrule
72 TeV & 10 & 2.53 (4.63) & 2.56 (4.68) & 2.63 (4.80) & 2.72 (4.95) \\
 & 30 & 1.94 (3.55) & 1.97 (3.60) & 2.04 (3.71) & 2.13 (3.87) \\
84 TeV & 10 & 2.31 (4.20) & 2.34 (4.24) & 2.42 (4.36) & 2.52 (4.51) \\
 & 30 & 1.77 (3.18) & 1.80 (3.22) & 1.88 (3.38) & 1.99 (3.57) \\
102 TeV & 10 & 2.12 (3.82) & 2.14 (3.86) & 2.20 (3.98) & 2.28 (4.13) \\
 & 30 & 1.60 (2.89) & 1.62 (2.93) & 1.69 (3.05) & 1.78 (3.22) \\
120 TeV & 10 & 1.88 (3.39) & 1.90 (3.43) & 1.96 (3.54) & 2.03 (3.69) \\
 & 30 & 1.42 (2.57) & 1.44 (2.61) & 1.49 (2.73) & 1.58 (2.88) \\
\bottomrule
\end{tabular}

%% file: q/tables/diss_limits.tex
\begin{tabular}{lcccc}
\toprule
$|\dtau|$ [$10^{-18}\,\ecm$], $10~\invab$ & 72~TeV & 84~TeV & 102~TeV & 120~TeV \\
\midrule
elastic only & 2.90 & 2.68 & 2.48 & 2.26 \\
A: dissociative as signal (baseline) & 2.63 & 2.42 & 2.20 & 1.96 \\
A with 50\% slope uncertainty & 2.80 & 2.56 & 2.31 & 2.04 \\
B: dissociative as SM-only background & 4.36 & 4.01 & 3.65 & 3.26 \\
\bottomrule
\end{tabular}

%% file: q/sections/conclusions.tex
\section{Conclusions}\label{sec:conclusions}

We have estimated the sensitivity of the FCC-hh to the anomalous magnetic moment and the electric dipole moment of
the $\tau$ lepton in photon-induced $\gaga\to\tau^+\tau^-$ production, including the central exclusive (elastic) and
the proton-dissociative contributions, using events in which both $\tau$ leptons
decay hadronically. The hadronic $\tau$ decays are reconstructed from particle-flow objects with an efficiency of
\res{reco-eff}\%, and after the exclusivity requirement the Drell--Yan and $\gaga\to WW$ backgrounds amount to
\res{y-bkg-rel-tot}\% of the SM signal in the fit range of the total transverse mass.

For the baseline scenario at $\sqrt{s} = 84$~TeV and $30~\invab$, the expected 95\% CL limits are
$\res{ataulo-84-30}\times10^{-4} < \atau < +\res{atauhi-84-30}\times10^{-4}$ and
$|\dtau| < \res{dtau-84-30}\times10^{-18}\,\ecm$, an improvement over the current best bounds by factors of about
\res{cms-a-ratio-84-30} and \res{cms-d-ratio-84-30}, respectively. The projected $\atau$ sensitivity is smaller than
the SM value itself. For the four FCC-hh operating scenarios between 72 and 120~TeV, ten years of operation at the nominal
luminosity yield $|\atau| \lesssim (\res{ten-a-best}$--$\res{ten-a-worst})\times10^{-4}$ and
$|\dtau| \lesssim (\res{ten-d-best}$--$\res{ten-d-worst})\times10^{-18}\,\ecm$. At equal luminosity the sensitivity
improves with the collider energy by \res{gain-10}\% from 72 to 120~TeV. This gain originates from the multi-TeV
region of the $\tau$-pair mass, as shown by restricting the anomalous contribution to lower masses.

Proton-dissociative production, estimated with \SuperChic\ and a generator-level evaluation of the exclusivity
requirement, contributes about twice as many events as the elastic process after the selection. Included as signal
with uncertainties on its normalisation and mass dependence, it improves the limits by
\res{diss-gain-min}--\res{diss-gain-max}\%. The comparison with \SuperChic\ further confirms the normalisation of the anomalous couplings and shows that the two descriptions of the elastic photon fluxes and of the
soft survival (\texttt{gamma}-UPC and \SuperChic) agree within a few per cent, changing the limits by less than
\res{K-eff-absmax}\%. The results depend only weakly on the assumed systematic uncertainty.

The main open issue is the exclusivity requirement in the presence of $\mathcal{O}(1000)$ pile-up interactions:
with the longitudinal track association used here nearly all signal events would be lost; a track-vertex association within a few hundred micrometers combined with precise timing information in the tracking detectors, and/or the tagging of the intact protons
in forward spectrometers, will therefore be required. Proton tagging would also suppress the proton-dissociative contributions. Since the simulation
keeps the $\tau$ spin correlations, CP-odd observables such as the angle between the $\tau$ decay planes can be
added to resolve the sign of $\dtau$ and separate $\atau$ from $\dtau$.

With its high centre-of-mass energy and integrated luminosity, together with advances in detector technology, the
FCC-hh will offer new opportunities to search for physics beyond the SM in the electromagnetic interactions of the
$\tau$ lepton. Photon-induced $\tau$-pair production provides a clean probe for this purpose since it involves only QED interactions at tree level. Its SM prediction depends mainly on the photon fluxes of the protons and the anomalous
contributions grow with the $\tau$-pair mass which reaches several TeV at the FCC-hh. The results presented here show
that this process can improve the sensitivity to the $\tau$ dipole moments by more than an order of magnitude with
respect to the current bounds.

%% file: q/sections/acknowledgements.tex
\section*{Acknowledgements}

This work was supported by the Turkish Energy, Nuclear and Mineral Research Agency (TENMAK), Grant
No.~2025TENMAK(CERN)A5.H3.F2-05. The numerical calculations reported in this paper were partially performed at
T\"UB\.ITAK ULAKB\.IM, High Performance and Grid Computing Center (TRUBA resources). The authors thank the CERN
Theoretical Physics Department for its hospitality during the visit by one of the authors where valuable
discussions related to this work took place.

%% file: refs.bib
@article{Harland-Lang:2020veo,
    author = "Harland-Lang, L. A. and Tasevsky, M. and Khoze, V. A. and Ryskin, M. G.",
    title = "{A new approach to modelling elastic and inelastic photon-initiated production at the LHC: SuperChic 4}",
    eprint = "2007.12704",
    archivePrefix = "arXiv",
    primaryClass = "hep-ph",
    reportNumber = "IPPP/20/33",
    doi = "10.1140/epjc/s10052-020-08455-0",
    journal = "Eur. Phys. J. C",
    volume = "80",
    number = "10",
    pages = "925",
    year = "2020"
}

@article{Harland-Lang:2024zpn,
    author = "Harland-Lang, L. A.",
    title = "{Higher precision constraints on the tau $g-2$ in LHC photon-initiated production: a full account of hadron dissociation and soft survival effects}",
    eprint = "2410.10978",
    archivePrefix = "arXiv",
    primaryClass = "hep-ph",
    doi = "10.1140/epjc/s10052-024-13685-7",
    journal = "Eur. Phys. J. C",
    volume = "84",
    number = "12",
    pages = "1332",
    year = "2024"
}

@article{CMS:2024skm,
    author = "{CMS Collaboration}",
    title = "{Observation of $\gamma\gamma\to\tau\tau$ in proton-proton collisions and limits on the anomalous electromagnetic moments of the $\tau$ lepton}",
    eprint = "2406.03975",
    archivePrefix = "arXiv",
    primaryClass = "hep-ex",
    reportNumber = "CMS-SMP-23-005, CERN-EP-2024-127",
    journal = "Rep. Prog. Phys.",
    volume = "87",
    pages = "107801",
    year = "2024"
}

@misc{CMS:HIN-24-011,
    author = "{CMS Collaboration}",
    title = "{Measurement of the $\gamma\gamma\to\tau^+\tau^-$ cross section and constraints on the anomalous magnetic moment of the $\tau$ lepton in ultraperipheral PbPb collisions at $\sqrt{s_{NN}}$ = 5.02 TeV}",
    eprint = "2607.20148",
    archivePrefix = "arXiv",
    primaryClass = "hep-ex",
    reportNumber = "CMS-HIN-24-011, CERN-EP-2026-160",
    howpublished = "\href{https://arxiv.org/abs/2607.20148}{arXiv:2607.20148 [hep-ex]}",
    note = "Submitted to Phys. Lett. B",
    year = "2026"
}

@article{Eidelman:2007sb,
    author = "Eidelman, S. and Passera, M.",
    title = "{Theory of the tau lepton anomalous magnetic moment}",
    eprint = "hep-ph/0701260",
    archivePrefix = "arXiv",
    doi = "10.1142/S0217732307022694",
    journal = "Mod. Phys. Lett. A",
    volume = "22",
    pages = "159--179",
    year = "2007"
}

@article{DELPHI:2003nah,
    author = "Abdallah, J. and others",
    collaboration = "DELPHI",
    title = "{Study of tau-pair production in photon-photon collisions at LEP and limits on the anomalous electromagnetic moments of the tau lepton}",
    eprint = "hep-ex/0406010",
    archivePrefix = "arXiv",
    reportNumber = "CERN-EP-2003-058",
    doi = "10.1140/epjc/s2004-01852-y",
    journal = "Eur. Phys. J. C",
    volume = "35",
    pages = "159--170",
    year = "2004"
}

@article{Belle:2021ybo,
    author = "Inami, K. and others",
    collaboration = "Belle",
    title = "{An improved search for the electric dipole moment of the $\tau$ lepton}",
    eprint = "2108.11543",
    archivePrefix = "arXiv",
    primaryClass = "hep-ex",
    reportNumber = "Belle Preprint 2021-22, KEK Preprint 2021-26",
    doi = "10.1007/JHEP04(2022)110",
    journal = "JHEP",
    volume = "04",
    pages = "110",
    year = "2022"
}

@article{ATLAS:2022ryk,
    author = "Aad, Georges and others",
    collaboration = "ATLAS",
    title = "{Observation of the {\ensuremath{\gamma}}{\ensuremath{\gamma}}{\textrightarrow}{\ensuremath{\tau}}{\ensuremath{\tau}} Process in Pb+Pb Collisions and Constraints on the {\ensuremath{\tau}}-Lepton Anomalous Magnetic Moment with the ATLAS Detector}",
    eprint = "2204.13478",
    archivePrefix = "arXiv",
    primaryClass = "hep-ex",
    reportNumber = "CERN-EP-2022-079",
    doi = "10.1103/PhysRevLett.131.151802",
    journal = "Phys. Rev. Lett.",
    volume = "131",
    number = "15",
    pages = "151802",
    year = "2023"
}

@article{CMS:2022arf,
    author = "Tumasyan, Armen and others",
    collaboration = "CMS",
    title = "{Observation of $\tau$ lepton pair production in ultraperipheral lead-lead collisions at $\sqrt{s_\mathrm{NN}}$ = 5.02 TeV}",
    eprint = "2206.05192",
    archivePrefix = "arXiv",
    primaryClass = "nucl-ex",
    reportNumber = "CMS-HIN-21-009, CERN-EP-2022-098",
    doi = "10.1103/PhysRevLett.131.151803",
    journal = "Phys. Rev. Lett.",
    volume = "131",
    pages = "151803",
    year = "2023"
}

@article{FCC:2018vvp,
    author = "Abada, A. and others",
    collaboration = "FCC",
    title = "{FCC-hh: The Hadron Collider}: {Future Circular Collider Conceptual Design Report Volume 3}",
    reportNumber = "CERN-ACC-2018-0058",
    doi = "10.1140/epjst/e2019-900087-0",
    journal = "Eur. Phys. J. ST",
    volume = "228",
    number = "4",
    pages = "755--1107",
    year = "2019"
}

@article{FCC:2025lpp,
    author = "Benedikt, M. and others",
    collaboration = "FCC",
    title = "{Future Circular Collider Feasibility Study Report: Volume 1, Physics, Experiments, Detectors}",
    eprint = "2505.00272",
    archivePrefix = "arXiv",
    primaryClass = "hep-ex",
    reportNumber = "CERN-FCC-PHYS-2025-0002",
    doi = "10.1140/epjc/s10052-025-15077-x",
    journal = "Eur. Phys. J. C",
    volume = "85",
    number = "12",
    pages = "1468",
    year = "2025",
    note = "[Erratum: Eur.Phys.J.C 86, 844 (2026)]"
}

@article{FCC:2025uan,
    author = "Benedikt, M. and others",
    collaboration = "FCC",
    title = "{Future Circular Collider Feasibility Study Report: Volume 2, Accelerators, Technical Infrastructure and Safety}",
    eprint = "2505.00274",
    archivePrefix = "arXiv",
    primaryClass = "physics.acc-ph",
    reportNumber = "CERN-FCC-ACC-2025-0004",
    doi = "10.1140/epjs/s11734-025-01967-4",
    journal = "Eur. Phys. J. ST",
    volume = "234",
    number = "19",
    pages = "5713--6197",
    year = "2025"
}

@article{Alwall:2014hca,
    author = "Alwall, J. and Frederix, R. and Frixione, S. and Hirschi, V. and Maltoni, F. and Mattelaer, O. and Shao, H. -S. and Stelzer, T. and Torrielli, P. and Zaro, M.",
    title = "{The automated computation of tree-level and next-to-leading order differential cross sections, and their matching to parton shower simulations}",
    eprint = "1405.0301",
    archivePrefix = "arXiv",
    primaryClass = "hep-ph",
    reportNumber = "CERN-PH-TH-2014-064, CP3-14-18, LPN14-066, MCNET-14-09, ZU-TH-14-14",
    doi = "10.1007/JHEP07(2014)079",
    journal = "JHEP",
    volume = "07",
    pages = "079",
    year = "2014"
}

@article{Bierlich:2022pfr,
    author = "Bierlich, Christian and others",
    title = "{A comprehensive guide to the physics and usage of PYTHIA 8.3}",
    eprint = "2203.11601",
    archivePrefix = "arXiv",
    primaryClass = "hep-ph",
    reportNumber = "LU-TP 22-16, MCNET-22-04, FERMILAB-PUB-22-227-SCD",
    doi = "10.21468/SciPostPhysCodeb.8",
    journal = "SciPost Phys. Codeb.",
    volume = "2022",
    pages = "8",
    year = "2022"
}

@article{deFavereau:2013fsa,
    author = "de Favereau, J. and Delaere, C. and Demin, P. and Giammanco, A. and Lema{\^\i}tre, V. and Mertens, A. and Selvaggi, M.",
    collaboration = "DELPHES 3",
    title = "{DELPHES 3, A modular framework for fast simulation of a generic collider experiment}",
    eprint = "1307.6346",
    archivePrefix = "arXiv",
    primaryClass = "hep-ex",
    doi = "10.1007/JHEP02(2014)057",
    journal = "JHEP",
    volume = "02",
    pages = "057",
    year = "2014"
}

@article{Hagiwara:2012vz,
    author = "Hagiwara, Kaoru and Li, Tong and Mawatari, Kentarou and Nakamura, Junya",
    title = "{TauDecay: a library to simulate polarized tau decays via FeynRules and MadGraph5}",
    eprint = "1212.6247",
    archivePrefix = "arXiv",
    primaryClass = "hep-ph",
    doi = "10.1140/epjc/s10052-013-2489-4",
    journal = "Eur. Phys. J. C",
    volume = "73",
    pages = "2489",
    year = "2013"
}

@article{CMS:2015pac,
    author = "Khachatryan, Vardan and others",
    collaboration = "CMS",
    title = "{Reconstruction and identification of $\tau$ lepton decays to hadrons and $\nu_\tau$ at CMS}",
    eprint = "1510.07488",
    archivePrefix = "arXiv",
    primaryClass = "physics.ins-det",
    reportNumber = "CMS-TAU-14-001, CERN-PH-EP-2015-261",
    doi = "10.1088/1748-0221/11/01/P01019",
    journal = "JINST",
    volume = "11",
    number = "01",
    pages = "P01019",
    year = "2016"
}

@article{CMS:2018jrd,
    author = "Sirunyan, A. M. and others",
    collaboration = "CMS",
    title = "{Performance of reconstruction and identification of $\tau$ leptons decaying to hadrons and $\nu_\tau$ in pp collisions at $\sqrt{s}=$ 13 TeV}",
    eprint = "1809.02816",
    archivePrefix = "arXiv",
    primaryClass = "hep-ex",
    reportNumber = "CMS-TAU-16-003, CERN-EP-2018-229",
    doi = "10.1088/1748-0221/13/10/P10005",
    journal = "JINST",
    volume = "13",
    number = "10",
    pages = "P10005",
    year = "2018"
}

@article{Beresford:2019gww,
    author = "Beresford, Lydia and Liu, Jesse",
    title = "{New physics and tau $g-2$ using LHC heavy ion collisions}",
    eprint = "1908.05180",
    archivePrefix = "arXiv",
    primaryClass = "hep-ph",
    doi = "10.1103/PhysRevD.102.113008",
    journal = "Phys. Rev. D",
    volume = "102",
    number = "11",
    pages = "113008",
    year = "2020",
    note = "[Erratum: Phys.Rev.D 106, 039902 (2022)]"
}

@article{Dyndal:2020yen,
    author = "Dyndal, Mateusz and Klusek-Gawenda, Mariola and Schott, Matthias and Szczurek, Antoni",
    title = "{Anomalous electromagnetic moments of $\tau$ lepton in $\gamma \gamma \to \tau^+ \tau^-$ reaction in Pb+Pb collisions at the LHC}",
    eprint = "2002.05503",
    archivePrefix = "arXiv",
    primaryClass = "hep-ph",
    doi = "10.1016/j.physletb.2020.135682",
    journal = "Phys. Lett. B",
    volume = "809",
    pages = "135682",
    year = "2020"
}

@misc{Inan:2026byu,
    author = "{\.I}nan, S. C. and Kisselev, A. V.",
    title = "{Probing electromagnetic moments of the tau lepton in PbPb collisions at the FCC-hh}",
    eprint = "2601.18288",
    archivePrefix = "arXiv",
    primaryClass = "hep-ph",
    howpublished = "\href{https://arxiv.org/abs/2601.18288}{arXiv:2601.18288 [hep-ph]}",
    year = "2026"
}

@article{Degrande:2011ua,
    author = "Degrande, Celine and Duhr, Claude and Fuks, Benjamin and Grellscheid, David and Mattelaer, Olivier and Reiter, Thomas",
    title = "{UFO - The Universal FeynRules Output}",
    eprint = "1108.2040",
    archivePrefix = "arXiv",
    primaryClass = "hep-ph",
    reportNumber = "CP3-11-25, IPHC-PHENO-11-04, IPPP-11-39, DCPT-11-78, MPP-2011-68",
    doi = "10.1016/j.cpc.2012.01.022",
    journal = "Comput. Phys. Commun.",
    volume = "183",
    pages = "1201--1214",
    year = "2012"
}

@article{Alloul:2013bka,
    author = "Alloul, Adam and Christensen, Neil D. and Degrande, C{\'e}line and Duhr, Claude and Fuks, Benjamin",
    title = "{FeynRules  2.0 - A complete toolbox for tree-level phenomenology}",
    eprint = "1310.1921",
    archivePrefix = "arXiv",
    primaryClass = "hep-ph",
    reportNumber = "CERN-PH-TH-2013-239, MCNET-13-14, IPPP-13-71, DCPT-13-142, PITT-PACC-1308",
    doi = "10.1016/j.cpc.2014.04.012",
    journal = "Comput. Phys. Commun.",
    volume = "185",
    pages = "2250--2300",
    year = "2014"
}

@article{Kuhn:1990ad,
    author = "Kuhn, Johann H. and Santamaria, A.",
    title = "{Tau decays to pions}",
    reportNumber = "MPI-PAE/PTh-17/90",
    doi = "10.1007/BF01572024",
    journal = "Z. Phys. C",
    volume = "48",
    pages = "445--452",
    year = "1990"
}

@article{FCC:2025jtd,
    author = "Benedikt, M. and others",
    collaboration = "FCC",
    title = "{Future Circular Collider Feasibility Study Report}",
    eprint = "2505.00273",
    archivePrefix = "arXiv",
    primaryClass = "physics.acc-ph",
    reportNumber = "CERN-FCC-ACC-2025-0003",
    doi = "10.1140/epjs/s11734-025-01958-5",
    journal = "Eur. Phys. J. ST",
    volume = "234",
    number = "17",
    pages = "5113--5383",
    year = "2025",
    note = "[Erratum: Eur.Phys.J.ST 234, 33--44 (2025)]"
}

@misc{Zimmermann:2024fcchh,
    author = "Zimmermann, F.",
    title = "{Scenarios for the FCC-hh}",
    howpublished = "{Note presented at the FCC-hh Studies for the next European Strategy kickoff meeting, CERN, 3 September 2024}",
    url = "https://indico.cern.ch/event/1439072/contributions/6106995/attachments/2917946/5120981/FCC_hh_scenarios.pdf",
    year = "2024"
}

@article{Sjostrand:2014zea,
    author = {Sj{\"o}strand, Torbj{\"o}rn and Ask, Stefan and Christiansen, Jesper R. and Corke, Richard and Desai, Nishita and Ilten, Philip and Mrenna, Stephen and Prestel, Stefan and Rasmussen, Christine O. and Skands, Peter Z.},
    title = "{An introduction to PYTHIA 8.2}",
    eprint = "1410.3012",
    archivePrefix = "arXiv",
    primaryClass = "hep-ph",
    reportNumber = "LU-TP-14-36, MCNET-14-22, CERN-PH-TH-2014-190, FERMILAB-PUB-14-316-CD, DESY-14-178, SLAC-PUB-16122",
    doi = "10.1016/j.cpc.2015.01.024",
    journal = "Comput. Phys. Commun.",
    volume = "191",
    pages = "159--177",
    year = "2015"
}

@article{Shao:2022cly,
    author = "Shao, Hua-Sheng and d'Enterria, David",
    title = "{$\gamma$-UPC: automated generation of exclusive photon-photon processes in ultraperipheral proton and nuclear collisions with varying form factors}",
    eprint = "2207.03012",
    archivePrefix = "arXiv",
    primaryClass = "hep-ph",
    doi = "10.1007/JHEP09(2022)248",
    journal = "JHEP",
    volume = "09",
    pages = "248",
    year = "2022"
}

@article{Crepet:2024dvv,
    author = "Cr{\'e}pet, Nicolas and d'Enterria, David and Shao, Hua-Sheng",
    title = "{Improved modeling of photon-photon processes in ultraperipheral collisions at hadron colliders}",
    eprint = "2409.18485",
    archivePrefix = "arXiv",
    primaryClass = "hep-ph",
    doi = "10.22323/1.469.0124",
    journal = "PoS",
    volume = "DIS2024",
    pages = "124",
    year = "2025"
}

@article{dEnterria:2019jty,
    author = "d'Enterria, David and Martins, Daniel E. and Rebello Teles, Patricia",
    title = "{Higgs boson production in photon-photon interactions with proton, light-ion, and heavy-ion beams at current and future colliders}",
    eprint = "1904.11936",
    archivePrefix = "arXiv",
    primaryClass = "hep-ph",
    doi = "10.1103/PhysRevD.101.033009",
    journal = "Phys. Rev. D",
    volume = "101",
    number = "3",
    pages = "033009",
    year = "2020"
}

@article{dEnterria:2025ecx,
    author = "d'Enterria, David and Kang, Karen",
    title = "{Exclusive photon-fusion production of even-spin resonances and exotic QED atoms in high-energy hadron collisions}",
    eprint = "2503.10952",
    archivePrefix = "arXiv",
    primaryClass = "hep-ph",
    doi = "10.1103/rnxl-v6gd",
    journal = "Phys. Rev. D",
    volume = "112",
    number = "11",
    pages = "116022",
    year = "2025"
}

@techreport{CMS:2014sdw,
    institution = "CERN",
    number = "CERN-LHCC-2014-021, TOTEM-TDR-003, CMS-TDR-013",
    author = "Albrow, M. and others",
    collaboration = "CMS, TOTEM",
    title = "{CMS-TOTEM Precision Proton Spectrometer}",
    reportNumber = "CERN-LHCC-2014-021, TOTEM-TDR-003, CMS-TDR-013, CERN-LHCC-2014-021, TOTEM-TDR-003, CMS-TDR-13",
    month = "9",
    year = "2014"
}

@techreport{Adamczyk:2015cjy,
    institution = "CERN",
    number = "CERN-LHCC-2015-009, ATLAS-TDR-024",
    author = "Adamczyk, L. and others",
    title = "{Technical Design Report for the ATLAS Forward Proton Detector}",
    reportNumber = "CERN-LHCC-2015-009, ATLAS-TDR-024",
    month = "5",
    year = "2015"
}

@article{CMS:2018uvs,
    author = "Sirunyan, Albert M and others",
    collaboration = "CMS, TOTEM",
    title = "{Observation of proton-tagged, central (semi)exclusive production of high-mass lepton pairs in pp collisions at 13 TeV with the CMS-TOTEM precision proton spectrometer}",
    eprint = "1803.04496",
    archivePrefix = "arXiv",
    primaryClass = "hep-ex",
    reportNumber = "CMS-PPS-17-001, TOTEM 2018-001, CERN-EP-2018-014",
    doi = "10.1007/JHEP07(2018)153",
    journal = "JHEP",
    volume = "07",
    pages = "153",
    year = "2018"
}

@article{ATLAS:2020mve,
    author = "Aad, Georges and others",
    collaboration = "ATLAS",
    title = "{Observation and Measurement of Forward Proton Scattering in Association with Lepton Pairs Produced via the Photon Fusion Mechanism at ATLAS}",
    eprint = "2009.14537",
    archivePrefix = "arXiv",
    primaryClass = "hep-ex",
    reportNumber = "CERN-EP-2020-167",
    doi = "10.1103/PhysRevLett.125.261801",
    journal = "Phys. Rev. Lett.",
    volume = "125",
    number = "26",
    pages = "261801",
    year = "2020"
}

@article{Atag:2010ja,
    author = "Atag, S. and Billur, A. A.",
    title = "{Possibility of Determining $\tau$ Lepton Electromagnetic Moments in ${\gamma\gamma \to \tau^{+}\tau^{-}}$ Process at the CERN-LHC}",
    eprint = "1005.2841",
    archivePrefix = "arXiv",
    primaryClass = "hep-ph",
    doi = "10.1007/JHEP11(2010)060",
    journal = "JHEP",
    volume = "11",
    pages = "060",
    year = "2010"
}

@article{Beresford:2024dsc,
    author = "Beresford, Lydia and Clawson, Savannah and Liu, Jesse",
    title = "{Strategy to measure tau g-2 via photon fusion in LHC proton collisions}",
    eprint = "2403.06336",
    archivePrefix = "arXiv",
    primaryClass = "hep-ph",
    doi = "10.1103/PhysRevD.110.092016",
    journal = "Phys. Rev. D",
    volume = "110",
    number = "9",
    pages = "092016",
    year = "2024"
}

@article{Billur:2013rva,
    author = "Billur, A. A. and Koksal, M.",
    title = "{Probe of the electromagnetic moments of the tau lepton in gamma-gamma collisions at the CLIC}",
    eprint = "1306.5620",
    archivePrefix = "arXiv",
    primaryClass = "hep-ph",
    doi = "10.1103/PhysRevD.89.037301",
    journal = "Phys. Rev. D",
    volume = "89",
    number = "3",
    pages = "037301",
    year = "2014"
}

@article{Koksal:2018env,
    author = {K{\"o}ksal, M. and Billur, A. A. and Guti{\'e}rrez-Rodr{\'\i}guez, A. and Hern{\'a}ndez-Ru{\'\i}z, M. A.},
    title = "{Model-independent sensitivity estimates for the electromagnetic dipole moments of the $\tau$-lepton at the CLIC}",
    eprint = "1804.02373",
    archivePrefix = "arXiv",
    primaryClass = "hep-ph",
    doi = "10.1103/PhysRevD.98.015017",
    journal = "Phys. Rev. D",
    volume = "98",
    number = "1",
    pages = "015017",
    year = "2018"
}

@article{Koksal:2018xyi,
    author = {K{\"o}ksal, M.},
    title = "{Search for the electromagnetic moments of the $\tau$ lepton in photon{\textendash}photon collisions at the LHeC and the FCC-he}",
    eprint = "1809.01963",
    archivePrefix = "arXiv",
    primaryClass = "hep-ph",
    doi = "10.1088/1361-6471/ab0b53",
    journal = "J. Phys. G",
    volume = "46",
    pages = "065003",
    year = "2019"
}

@article{Denizli:2024uwv,
    author = {Denizli, H. and Senol, A. and K{\"o}ksal, M.},
    title = "{Constraints on \(\tau\) electromagnetic moments via tau pair production at the Muon colliders}",
    eprint = "2408.16106",
    archivePrefix = "arXiv",
    primaryClass = "hep-ph",
    doi = "10.1016/j.cjph.2025.04.020",
    journal = "Chin. J. Phys.",
    volume = "95",
    pages = "1250--1258",
    year = "2025"
}

@article{Shao:2025xwp,
    author = "Shao, Ding Yu and Xiang, Hao and Xu, Fang and Yan, Bin and Zhang, Cheng",
    title = "{Linearly polarized photon fusion as a precision probe of the tau lepton dipole moments at lepton colliders}",
    eprint = "2506.15245",
    archivePrefix = "arXiv",
    primaryClass = "hep-ph",
    doi = "10.1103/l2j6-mbwx",
    journal = "Phys. Rev. D",
    volume = "113",
    number = "3",
    pages = "035018",
    year = "2026"
}

@misc{Bhide:2024nje,
    author = "Bhide, Kartik and Lang, Valerie",
    title = "{Probing optimal measurements of the electromagnetic dipole moments of the $\tau$ lepton}",
    eprint = "2410.23070",
    howpublished = "\href{https://arxiv.org/abs/2410.23070}{arXiv:2410.23070 [hep-ph]}",
    archivePrefix = "arXiv",
    primaryClass = "hep-ph",
    year = "2024"
}

@article{Koksal:2017nmy,
    author = {K{\"o}ksal, M. and {\.I}nan, S. C. and Billur, A. A. and {\"O}zg{\"u}ven, Y. and Bahar, M. K.},
    title = "{Analysis of the anomalous electromagnetic moments of the tau lepton in $\gamma p$ collisions at the LHC}",
    eprint = "1711.02405",
    archivePrefix = "arXiv",
    primaryClass = "hep-ph",
    doi = "10.1016/j.physletb.2018.07.018",
    journal = "Phys. Lett. B",
    volume = "783",
    pages = "375--380",
    year = "2018"
}

@article{Fomin:2018ybj,
    author = "Fomin, A. S. and Korchin, A. Yu and Stocchi, A. and Barsuk, S. and Robbe, P.",
    title = "{Feasibility of $\tau$ -lepton electromagnetic dipole moments measurement using bent crystal at the LHC}",
    eprint = "1810.06699",
    archivePrefix = "arXiv",
    primaryClass = "hep-ph",
    doi = "10.1007/JHEP03(2019)156",
    journal = "JHEP",
    volume = "03",
    pages = "156",
    year = "2019"
}

@article{Chen:2018cxt,
    author = "Chen, Xin and Wu, Yongcheng",
    title = "{Search for the Electric Dipole Moment and anomalous magnetic moment of the tau lepton at tau factories}",
    eprint = "1803.00501",
    archivePrefix = "arXiv",
    primaryClass = "hep-ph",
    doi = "10.1007/JHEP10(2019)089",
    journal = "JHEP",
    volume = "10",
    pages = "089",
    year = "2019"
}

@article{Gutierrez-Rodriguez:2022eyn,
    author = {Guti{\'e}rrez-Rodr{\'\i}guez, A. and K{\"o}ksal, M. and Billur, A. A. and Hern{\'a}ndez-Ru{\'\i}z, M. A.},
    title = "{Feasibility at the LHC, FCC-he and CLIC for Sensitivity Estimates on Anomalous Electromagnetic Couplings of the $\tau$-lepton}",
    eprint = "1903.04135",
    archivePrefix = "arXiv",
    primaryClass = "hep-ph",
    doi = "10.1007/s10773-022-05150-6",
    journal = "Int. J. Theor. Phys.",
    volume = "61",
    number = "6",
    pages = "161",
    year = "2022"
}

@article{Ball:2012cx,
    author = "Ball, Richard D. and others",
    title = "{Parton distributions with LHC data}",
    eprint = "1207.1303",
    archivePrefix = "arXiv",
    primaryClass = "hep-ph",
    reportNumber = "EDINBURGH-2012-08, IFUM-FT-997, FR-PHENO-2012-014, RWTH-TTK-12-25, CERN-PH-TH-2012-037, SFB-CPP-12-47",
    doi = "10.1016/j.nuclphysb.2012.10.003",
    journal = "Nucl. Phys. B",
    volume = "867",
    pages = "244--289",
    year = "2013"
}

@article{Baker:1983tu,
    author = "Baker, Steve and Cousins, Robert D.",
    title = "{Clarification of the Use of Chi Square and Likelihood Functions in Fits to Histograms}",
    reportNumber = "UCLA-HEP-83-3",
    doi = "10.1016/0167-5087(84)90016-4",
    journal = "Nucl. Instrum. Meth.",
    volume = "221",
    pages = "437--442",
    year = "1984"
}

@article{Cowan:2010js,
    author = "Cowan, Glen and Cranmer, Kyle and Gross, Eilam and Vitells, Ofer",
    title = "{Asymptotic formulae for likelihood-based tests of new physics}",
    eprint = "1007.1727",
    archivePrefix = "arXiv",
    primaryClass = "physics.data-an",
    doi = "10.1140/epjc/s10052-011-1554-0",
    journal = "Eur. Phys. J. C",
    volume = "71",
    pages = "1554",
    year = "2011",
    note = "[Erratum: Eur.Phys.J.C 73, 2501 (2013)]"
}
